\documentclass[conference]{IEEEtran}
\IEEEoverridecommandlockouts
\usepackage{cite}
\usepackage{amsmath,amssymb,amsfonts}
\usepackage{algorithmic}
\usepackage{graphicx}
\usepackage{textcomp}
\usepackage{xcolor}
\def\BibTeX{{\rm B\kern-.05em{\sc i\kern-.025em b}\kern-.08em
    T\kern-.1667em\lower.7ex\hbox{E}\kern-.125emX}}

\usepackage{amsmath,amssymb,amsfonts}
\usepackage{algorithmic}
\usepackage{booktabs}
\usepackage{graphicx}
\usepackage{textcomp, multirow}
\usepackage{bmpsize}
\PassOptionsToPackage{hyphens}{url}\usepackage{hyperref}
\usepackage{breakcites}\usepackage{xcolor}
\usepackage{lipsum}
\usepackage{amsmath}      
\usepackage{algorithm}    
\usepackage{algorithmic}  
\usepackage{colortbl}
\usepackage{xcolor}
\usepackage{pgf}  

\newcommand{\deltaValue}[1]{%
    \pgfmathparse{#1 > 5 ? 1 : 0} 
    \ifnum\pgfmathresult=1
        \cellcolor{blue!50}#1 
    \else
        \pgfmathparse{#1 < -5 ? 1 : 0} 
        \ifnum\pgfmathresult=1
            \cellcolor{red!50}#1 
        \else
            #1 
        \fi
    \fi
}

\newcommand{\colorValue}[1]{%
    \pgfmathparse{#1<50 ? 1 : 0} 
    \ifnum\pgfmathresult=1
        \cellcolor{red!80} #1 
    \else
        \pgfmathsetmacro{\percent}{(#1 - 50)/50}  
        \pgfmathsetmacro{\red}{255*(1-\percent)}   
        \pgfmathsetmacro{\green}{255*\percent}     
        \pgfmathsetmacro{\redInt}{int(\red)}       
        \pgfmathsetmacro{\greenInt}{int(\green)}   
        \xdef\colorCode{\noexpand\cellcolor[RGB]{\redInt,\greenInt,0}}%
        \colorCode #1
    \fi
}

\usepackage{hyperref}
\def\BibTeX{{\rm B\kern-.05em{\sc i\kern-.025em b}\kern-.08em
    T\kern-.1667em\lower.7ex\hbox{E}\kern-.125emX}}
\usepackage{cleveref}
\begin{document}

\title{DeePen: Penetration Testing for Audio Deepfake Detection}

\makeatletter
\newcommand{\linebreakand}{%
  \end{@IEEEauthorhalign}
  \hfill\mbox{}\par
  \mbox{}\hfill\begin{@IEEEauthorhalign}
}
\makeatother


\author{\IEEEauthorblockN{{ Nicolas M. M\"uller
}}
\IEEEauthorblockA{\textit{Fraunhofer AISEC}\\
Garching near Munich, Germany \\
nicolas.mueller@aisec.fraunhofer.de}
\and
\IEEEauthorblockN{Piotr Kawa}
\IEEEauthorblockA{
\textit{Wroc{\l}aw University}\\
\textit{of Science and Technology}\\
\textit{Resemble AI}\\
Wroc{\l}aw, Poland 
Mountain View, CA, USA \\
piotr.kawa@pwr.edu.pl}
\and
\IEEEauthorblockN{Adriana Stan}
\IEEEauthorblockA{
\textit{Technical University of Cluj-Napocay}\\
Cluj-Napoca, Romania \\
adriana.stan@com.utcluj.ro}
\linebreakand
\IEEEauthorblockN{Thien-Phuc Doan, Souhwan Jung}
\IEEEauthorblockA{
\textit{AISRC, Soongsil University}\\
Seoul, South Korea \\
phucdt@soongsil.ac.kr, souhwanj@ssu.ac.kr}
\and

\IEEEauthorblockN{Wei Herng Choong, Philip Sperl, Konstantin Böttinger}
\IEEEauthorblockA{\textit{Fraunhofer AISEC}\\
Garching near Munich, Germany \\
\{firstname.lastname\}@aisec.fraunhofer.de}
}

\maketitle

\begin{abstract}



Deepfakes—manipulated or forged audio and video media—pose significant security risks to individuals, organizations, and society at large. To address these challenges, machine-learning-based classifiers are commonly employed to detect deepfake content.

In this paper, we assess the robustness of such classifiers through a systematic penetration testing methodology, which we introduce as \emph{DeePen}. Our approach operates without prior knowledge of or access to the target deepfake detection models. Instead, it leverages a set of carefully selected signal processing modifications—referred to as attacks—to evaluate model vulnerabilities.

Using \emph{DeePen}, we analyze both real-world production systems and publicly available academic model checkpoints, demonstrating that all tested systems exhibit weaknesses and can be reliably deceived by simple manipulations such as time-stretching or echo addition. Furthermore, our findings reveal that while some attacks can be mitigated by retraining detection systems with knowledge of the specific attack, others remain persistently effective. 

\end{abstract}


\section{Introduction}

Generative AI has made remarkable strides across all modalities, including text, images, and audio. In the realm of audio, text-to-speech (TTS) technology unlocks a wide range of practical applications. For instance, in the television and film industry, TTS can efficiently generate audio in diverse voices, benefiting smaller studios and enabling the recreation of legacy characters' voices. In healthcare, TTS can restore speech capabilities for individuals with disabilities, as exemplified by projects like Google's Parrotron~\cite{parrotron} and others~\cite{personal-voice}.

However, TTS technology also presents significant risks due to its potential for misuse. Deepfakes, or \emph{spoofs}, are artificially generated voice recordings that can falsely attribute speech to individuals. This technology has been exploited to spread disinformation, erode trust in government authorities, and defame public figures and private individuals alike~\cite{deep_misinfo,df_taylor,dfporn}. Furthermore, deepfakes have been used in fraudulent schemes, phishing attacks, and to steal sensitive information~\cite{deepfake_fraud,deepfake_fraud2,deepfake_stock}, and have even been deployed in conventional state-on-state warfare~\cite{zel_fake}.
The solution is to build accurate deepfake detection systems to determine whether a given audio sample is genuine (or \emph{bona-fide}) or fabricated (or \emph{spoofed}). 
This technology could play a crucial role in combating misinformation by detecting manipulated media, supporting fact-checking efforts, and providing valuable evidence in legal contexts.

Given the high stakes, ensuring the robustness of deepfake detection systems is essential. However, in this paper, we show that even state-of-the-art commercial and open source deepfake detectors can be bypassed by making slight alterations to the audio sample.
These techniques enable attackers to reliably bypass detection, thereby compromising the credibility of these systems. Specifically, our contributions are as follows:

(i) We introduce an \textbf{open-source tool, \emph{DeePen}} for evaluating the robustness of audio deepfake detection models; 

(ii) We \textbf{evaluate three commercial and six open-source models using DeePen}, and identify their vulnerabilities. Notably, we show that it is relatively straightforward for an attacker to disguise deepfakes as authentic audio; 

(iii) We evaluate DeePen in the context of an \textbf{adaptive defender} who is aware of the model's vulnerabilities. 
We show that adaptive retraining can mitigate some, but not all effects discovered using the DeePen attacks;

(iv) Finally, we demonstrate that a subset of \textbf{adaptive defensive augmentations} achieves comparable performance to the naive approach (i.e., indiscriminately applying all defensive augmentations). This subset generalizes to previously unseen attacks and can, therefore, be regarded as a ``minimal representative set'' of defenses. This finding offers insights into the specific audio features that the deepfake detection system learns.

\section{Related Work}


Audio deepfake detection, or anti-spoofing, has seen an impressive increase in the number of scientific publications and commercial tool releases over the last few years. There are dedicated challenges (e.g. ASVspoof~\cite{asvpaper24} or ADD~\cite{add2023}) and special sessions at the major conferences (e.g., Interspeech or ICASSP). These challenges and sessions reveal the growing interest in deepfake countermeasures, which is strongly correlated to the threats to cybersecurity that deepfake generation methods pose. Audio deepfake generation is performed by Text-to-Speech (TTS) or Voice Conversion (VC) methods, and systems such as 
ElevenLabs, Resemble AI or Respeecher  can now copy a target speaker's voice with only a few seconds of audio~\cite{eleven-labs,respeecher,resemble-ai}.

Some of the early stages of deepfake detection systems have relied on building dedicated machine learning architectures that process raw waveform~\cite{rawnet, aasist,raw-pc-darts}, spectral features~\cite{9023158,wu12c_interspeech,lcnn} or learnable deep features~\cite{tomilov21_asvspoof,rawnet-2}.
However, with the advent of large pre-trained self-supervised learning (SSL) audio models (e.g. wav2vec 2.0,  wavLM, Whisper and their variants), the deepfake detection methods have shifted focus and exploit the inherent audio representations that these models learn.
The SSL models are used as feature extractors (or \textit{frontends}), and simple~\cite{pascu24_interspeech,whisper-df} or more complex~\cite{wu24c_interspeech} classifier modules (or \textit{backends}) are appended to the pre-trained~\cite{saha24green} or finetuned~\cite{wav2vec_aasist} SSL component.

One problem that most of these systems encounter is the fact that they do not properly generalise to out-of-distribution samples -- meaning that they do not learn meaningful audio attributes, but rather the classification is performed by mostly
exploiting potential data processing oversights in the available samples. 
For example, M\"uller et al.~\cite{muller2022interspeech} and Zhang et al.~\cite{10224301} showed that the length of silence segments can be an accurate indicator of fake versus real data for some datasets. In a separate study M\"uller et al.~\cite{Mller2021SpeechIS} determined a correlation between the length of the audio sample and the final decision, 
while Negroni et al.~\cite{negroni2024analyzing} exploited low frequency artefacts at splicing points in the audio signal for partial deepfake detection.

To overcome the lack of generalisation, 
data augmentation~\cite{10557138} can be set in place. The audio training set can be extended with real~\cite{lu24b_interspeech}, vocoded~\cite{vocoded_df} or synthesised samples~\cite{martindonas24_interspeech}; or by simply performing signal-based modifications to the existing samples (e.g. adding noise, reverberation, truncating the signal, scrambling, etc.)~\cite{rawboost,tak2022automatic}. The major issue of data augmentation is the proper selection of additional samples and signal processing methods which are informative for the final decision, which is not a trivial task~\cite{azeemi23_interspeech,10448016}.

To make things even more complex, a very recent security threat is that of adversarial attacks targeting the deepfake detection systems themselves~\cite{panariello23b_interspeech, advshadow, malacopula2024}. The attacks imperceptibly alter a generated audio sample to fool a particular or several deepfake detectors. 





\section{DeePen Methodology} 
\begin{figure*}[t]
    \centering
    \includegraphics[width=0.60\linewidth]{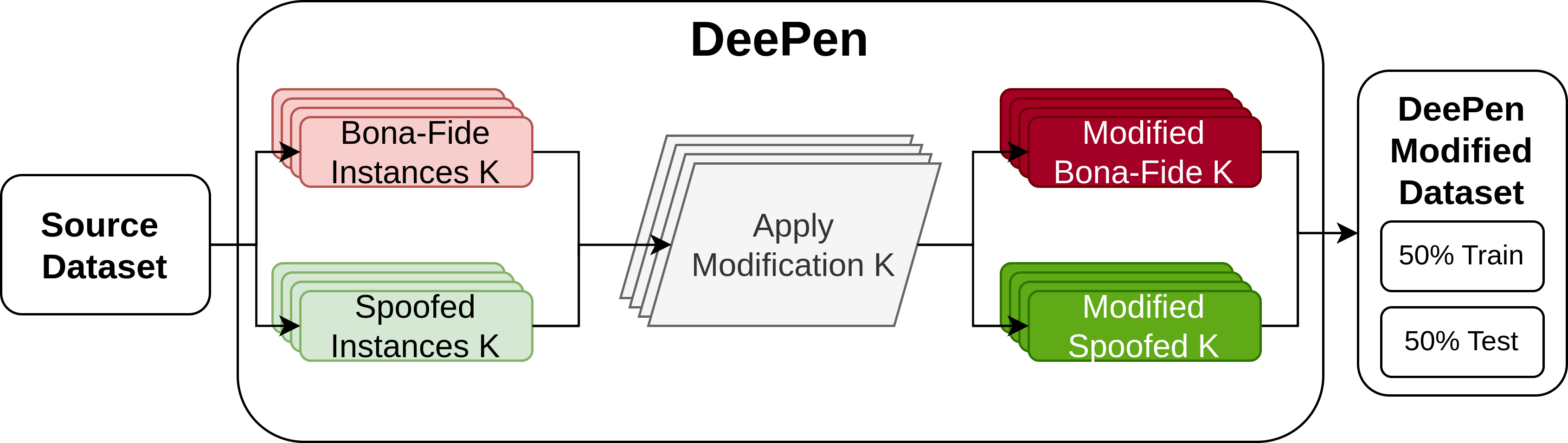}
    \caption{Application of DeePen methodology to an existing anti-spoofing dataset.
    Given a dataset such as ASVspoof 2019 or MLAAD, we extract $N$ random files of bona-fide and spoofed data, respectively. Each sample is modified by an individual attack $k$ out of the $K=17$ listed in \Cref{tab:audio_modifications}. The resulting $N \times 2 \times K$ shards of data constitute the DeePen methodology-derived dataset.}
    \label{fig:meth}
\end{figure*}

We present an approach for building a penetration test dataset starting from any available deepfake detection set of samples. An overview of our method is shown in~\Cref{fig:meth}.
Given a source dataset, we first select $N$ random samples from the bona-fide and spoofed classes, respectively. For each of these samples we apply one of the 
$K=17$ audio modifications listed in \Cref{tab:audio_modifications}. We refer to these modifications as \emph{attacks}, and their 
individual parameters are randomly set at generation time.
The resulting samples are then split into \emph{train} and \emph{test} partitions. The train partition can be used to augment the original dataset, while the test set can provide the final robustness evaluation for the novel attacks. 

In our experiments, we start from two established audio deepfake datasets.
First, we use ASVspoof 2019~\cite{asvspoof-2019}, which is the canonical benchmark for training and evaluating anti-spoofing detection systems~\cite{raw-pc-darts,rawnet-2,wav2vec_aasist,resnet-df,in-the-wild,aad,rawgat-st, complex-net,ockd,samo}. 
The dataset includes 18 different Text-to-Speech (TTS) systems, alongside a collection of corresponding genuine audio samples. 
Second, we employ the Multi-Language Audio Anti-Spoofing Dataset (MLAAD)~\cite{mlaad}. 
MLAAD encompasses some of the more recent audio deepfake generation methods for over 25 languages. There are more than 55 state-of-the-art TTS systems, including Microsoft T5~\cite{speech-t5}, XTTS~\cite{xtts}, Tortoise~\cite{tortoise}, among others.

From each of the two datasets we select $N=200$ random samples and generate a final penetration set of $2\times7200$ audio samples. 
These are then partitioned into a ``train'' and ``test'' set, with an equal 50/50 split.

\begin{table}[t]
\centering

\begin{tabular}{|p{7.4cm}|}
\hline

\textbf{No Attack}: Original audio, no modifications.\\ \hline
\textbf{Add Background Music}: Music from \texttt{Musan}~\cite{musan} or \texttt{Free Music Archive}~\cite{free-music-archive} overlaid at 50\% volume. \\ \hline
\textbf{Add Background Noise}: Noise from \texttt{Noise ESC 50}~\cite{esc-50} or \texttt{Musan}~\cite{musan} datasets added at 50\% relative volume. \\ \hline
\textbf{Amplitude Modulation}: Amplitude modulated by multiplying the signal with a sine wave of frequency between 0.5 and 5 Hz. \\ \hline
\textbf{Autotune}: Pitch corrected to the nearest value in a predefined scale (e.g., C major) using~\cite{autotune-software}. \\ \hline
\textbf{Bit Depth Change}: Reduced to 8 bits using \texttt{pydub.AudioSegment.set\_sample\_width(8)}~\cite{pydub}. \\ \hline
\textbf{Echo}: Echo added with delay between 0.1 and 1 second and decay between 0.3 and 0.9. \\ \hline
\textbf{Equalization}: 2 to 10 bands (1,000–10,000 Hz) with gains between ±4 and 15 dB. \\ \hline
\textbf{Freq Minus}: Subtracts 0.01–0.1 energy at random STFT frequencies between 0 and 4300 Hz. \\ \hline
\textbf{Freq Plus}: Adds 0.01–0.1 energy at random STFT frequencies between 0 and 4300 Hz. \\ \hline
\textbf{Gaussian Noise}: Gaussian noise with mean 0 and a standard deviation randomly selected between 0.01 and 0.2 added to the audio. \\ \hline
\textbf{High Pass Filter}: Butterworth high-pass filter applied with a cutoff frequency randomly selected between 2000 and 4000 Hz. \\ \hline
\textbf{Low Pass Filter}: Butterworth low-pass filter applied with a cutoff frequency randomly selected between 300 and 3000 Hz. \\ \hline
\textbf{MP3 Compression}: Converted to MP3 with a randomly selected bitrate between 4 and 48 kbps using \texttt{pydub.AudioSegment.export()}~\cite{pydub}. \\ \hline
\textbf{Pitch Shift}: Pitch shifted by -5 to +5 semitones using \texttt{librosa.effects.pitch\_shift()}~\cite{librosa}. \\ \hline
\textbf{Reverb}: Reverb applied via convolution with a randomly chosen decay factor between 1 and 10. \\ \hline
\textbf{Silence Injection}: Inserts a silent segment of random length between 0.1 and 2 seconds at the beginning of the audio. \\ \hline
\textbf{Time Stretch}: Audio speed modified by a factor between 0.8 (slower) and 1.2 (faster) using \texttt{librosa.effects.time\_stretch()}~\cite{librosa}. \\ \hline
\end{tabular}

\vspace{0.1cm}
\caption{DeePen attacks and their parameters}
\label{tab:audio_modifications}
\end{table}

\section{Evaluation} 
To evaluate DeePen, 
we define a three-step evaluation procedure as follows.

\begin{enumerate}
    \item \emph{Open-source deepfake detection models}. Here, we evaluate DeePen's influence over six academically released checkpoints.
    \item \emph{Commercial deepfake detection models}. This evaluates DeePen's influence over four commercial audio deepfake detection systems.
    \item \emph{Adaptive defense}. Here, we utilize the attacks from DeePen as a ``defense'' strategy and re-train the defender's model on them. We then evaluate whether this can mitigate the attacks.
\end{enumerate}
We always provide a \emph{no attack} baseline to assess the systems' de-facto effectiveness, i.e. its performance when not under attack.

\subsection{Open-source deepfake detection models}
\label{ss:open-models}

The first step in the evaluation of the DeePen methodology is to test it against a selection of readily available systems from open-source platforms. 
The models and trained checkpoints are provided by their respective authors. 
These models are ready to be deployed and serve as an estimate of the models used by the research community. We select the following six systems:

\begin{itemize}
    \item \textbf{Raw PC-DARTS}\footnote{\url{https://github.com/eurecom-asp/raw-pc-darts-anti-spoofing}}~\cite{raw-pc-darts} is a raw waveform processing architecture. It uses a differentiable architecture search (DARTS)~\cite{darts} process to determine the preprocessing steps and the exact look of the architecture's components. The architecture starts with sinc filters~\cite{sinc-net}, followed by the learnable blocks (called ``cells''), and ends with a recurrent GRU module~\cite{gru} with a fully connected layer.
    \item \textbf{LCNN}\footnote{\url{https://github.com/asvspoof-challenge/2021}}~\cite{lcnn} is a lightweight recurrent convolutional model. The convolution blocks included in the model, instead of a standard activation function such as ReLU, are based on Max-Feature-Map layers to improve feature selection. The used LCNN implementation processes audio transformed into linear frequency cepstral coefficient (LFCC) features~\cite{lfcc-vs-mfcc}. 
    \item \textbf{RawGAT-ST}\footnote{\url{https://github.com/eurecom-asp/RawGAT-ST-antispoofing}}~\cite{rawgat-st}, is a raw audio processing spectro-temporal graph attention network (GAT). The model learns the relationships between artifacts hidden in different sub-bands and temporal segments of audio. RawGAT-ST is composed of 3 GAT layers --- spectral and temporal ones and the spectro-temporal which processes the fusion of their outputs.

    \item \textbf{RawNet2}\footnote{\url{https://github.com/asvspoof-challenge/2021}}~\cite{rawnet-2} is a recurrent raw waveform architecture, originally adapted from the speaker recognition field. It is based on a~SincConv module~\cite{sinc-net}, followed by the groups of residual blocks with their output processed by the GRU layer~\cite{gru}.
    \item \textbf{WhisperDF}\footnote{\url{https://github.com/piotrkawa/deepfake-whisper-features}}~\cite{whisper-df} is an architecture based on the features obtained using the pre-trained Whisper automatic speech recognition (ASR) model~\cite{whisper}. The used implementation is based on the backbone of MesoNet convolutional network~\cite{mesonet} processing input of Whisper features concatenated with the mel-frequency cepstral coefficients (MFCC)~\cite{lfcc-vs-mfcc}.
    \item \textbf{W2V2}\footnote{\url{https://github.com/TakHemlata/SSL_Anti-spoofing}}~\cite{wav2vec_aasist} builds upon the AASIST model~\cite{aasist}, which leverages Sinc-Convolutions to extract features from raw audio inputs and incorporates a graph attention mechanism as its backend. 
    In contrast, W2V2 employs a Wav2Vec2.0~\cite{wav2vec} self-supervised learning (SSL) feature extractor, consisting of 300 million parameters. Furthermore, the model training utilized an innovative data augmentation approach known as RawBoost~\cite{rawboost}.

\end{itemize}

\Cref{tab:github} presents the classification results of the open-source models when the DeePen penetration test is performed. The columns represent the different datasets (ASVspoof 2019 and MLAAD), as well as the six deepfake detection models. 
The rows correspond to the DeePen type of attack for each class of samples, i.e. bona-fide or spoof.
For instance, the value of $22.0$ in the top-left corner and second line indicates that the Darts deepfake detection model has an average accuracy of $22.0$\% when bona-fide data from ASVspoof 2019 has been modified using the ``Add Background Music'' attack. 
This suggests that this attack breaks the Darts model, causing it to misclassify most bona-fide instances as spoof due to the presence of background music.
Notice that in the ``no attack'' case shown in the first line, Darts performs to $99\%$ accuracy for bona-fide data from ASVspoof 2019; thus the DeePen attack has caused an accuracy drop of $77\%$ percentage points.
The overall trends we observe in \Cref{tab:github} are related to:


\begin{table*}[ht]
    \centering
    \resizebox{.99\textwidth}{!}{
    \begin{tabular}{|l|l|p{0.7cm}p{0.7cm}p{0.9cm}p{0.9cm}p{0.9cm}p{0.7cm}|p{0.7cm}p{0.7cm}p{0.9cm}p{0.9cm}p{0.9cm}p{0.7cm}|}
\hline
    & &\multicolumn{12}{c|}{\textbf{Accuracy [\%] $\uparrow$}} \\ \cline{3-14}
\multirow{2}{*}{Label} & \multirow{2}{*}{Attack} & \multicolumn{6}{c|}{\textbf{ASVspoof 2019}}       & \multicolumn{6}{c|}{\textbf{MLAAD}}            \\ \cline{3-14}
                         &     & DARTS & LCNN & RawGAT & RawNet2 & Whisper & W2V2 & DARTS & LCNN & RawGAT & RawNet2 & Whisper & W2V2 \\ 
                         \hline \hline
\multirow{18}{*}{\rotatebox{90}{BONA-FIDE}}  
 & \textbf{No Attack}                & \colorValue{99.0}  & \colorValue{99.0}  & \colorValue{100.0} & \colorValue{100.0} &\colorValue{92.0} & \colorValue{100.0}  & \colorValue{19.0} & \colorValue{0.0} & \colorValue{95.0} & \colorValue{96.0}  &\colorValue{77.0} & \colorValue{93.0}  \\  \cline{2-14}
 & Add Background Music & \colorValue{22.0}  & \colorValue{39.0}  & \colorValue{27.0}  & \colorValue{64.0}  &\colorValue{34.0} & \colorValue{64.0}  & \colorValue{2.0}  & \colorValue{26.0} & \colorValue{22.0} & \colorValue{57.0}  &\colorValue{32.0} & \colorValue{50.0}  \\ \cline{2-14}
 & Add Background Noise & \colorValue{47.0}  & \colorValue{60.0}  & \colorValue{52.0}  & \colorValue{70.0}  &\colorValue{48.0} & \colorValue{93.0}  & \colorValue{10.0} & \colorValue{38.0} & \colorValue{29.0} & \colorValue{49.0}  &\colorValue{55.0} & \colorValue{66.0}  \\ \cline{2-14}
 & Amplitude Modulation & \colorValue{97.0}  & \colorValue{94.0}  & \colorValue{99.0}  & \colorValue{87.0}  &\colorValue{19.0} & \colorValue{75.0}  & \colorValue{28.0} & \colorValue{0.0} & \colorValue{95.0} & \colorValue{89.0}  &\colorValue{9.0} & \colorValue{14.0}  \\ \cline{2-14}
 & Autotune             & \colorValue{98.0}  & \colorValue{96.0}  & \colorValue{99.0}  & \colorValue{96.0}  &\colorValue{72.0} & \colorValue{93.0}  & \colorValue{21.0} & \colorValue{0.0} & \colorValue{92.0} & \colorValue{95.0}  &\colorValue{32.0} & \colorValue{65.0}  \\ \cline{2-14}
 & Bit Depth Change     & \colorValue{98.0}  & \colorValue{39.0}  & \colorValue{99.0}  & \colorValue{97.0}  &\colorValue{57.0} & \colorValue{99.0}  & \colorValue{16.0} & \colorValue{0.0} & \colorValue{90.0} & \colorValue{92.0}  &\colorValue{41.0} & \colorValue{84.0}  \\ \cline{2-14}
 & Echo                 & \colorValue{76.0}  & \colorValue{85.0}  & \colorValue{77.0}  & \colorValue{79.0}  &\colorValue{21.0} & \colorValue{46.0}  & \colorValue{13.0} & \colorValue{0.0} & \colorValue{44.0} & \colorValue{60.0}  &\colorValue{0.0} & \colorValue{2.0}  \\ \cline{2-14}
 & Equalization         & \colorValue{99.0}  & \colorValue{98.0}  & \colorValue{99.0}  & \colorValue{98.0}  &\colorValue{93.0} & \colorValue{100.0}  & \colorValue{25.0} & \colorValue{2.0} & \colorValue{87.0} & \colorValue{91.0}  &\colorValue{74.0} & \colorValue{84.0}  \\ \cline{2-14}
 & Freq Minus           & \colorValue{41.0}  & \colorValue{12.0}  & \colorValue{37.0}  & \colorValue{83.0}  &\colorValue{92.0} & \colorValue{100.0}  & \colorValue{4.0}  & \colorValue{0.0} & \colorValue{27.0} & \colorValue{77.0}  &\colorValue{85.0} & \colorValue{94.0}  \\ \cline{2-14}
 & Freq Plus            & \colorValue{44.0}  & \colorValue{18.0}  & \colorValue{39.0}  & \colorValue{84.0}  &\colorValue{92.0} & \colorValue{99.0}  & \colorValue{9.0}  & \colorValue{0.0} & \colorValue{27.0} & \colorValue{77.0}  &\colorValue{81.0} & \colorValue{96.0}  \\ \cline{2-14}
 & Gaussian Noise       & \colorValue{18.0}  & \colorValue{43.0}  & \colorValue{11.0}  & \colorValue{100.0} &\colorValue{21.0} & \colorValue{99.0}  & \colorValue{1.0}  & \colorValue{79.0} & \colorValue{34.0} & \colorValue{99.0}  &\colorValue{5.0} & \colorValue{74.0}  \\ \cline{2-14}
 & High Pass Filter     & \colorValue{99.0}  & \colorValue{95.0}  & \colorValue{77.0}  & \colorValue{99.0}  & \colorValue{29.0} & \colorValue{86.0}  & \colorValue{99.0} & \colorValue{11.0} & \colorValue{43.0} & \colorValue{100.0} & \colorValue{28.0} & \colorValue{53.0}  \\ \cline{2-14}
 & Low Pass Filter      & \colorValue{98.0}  & \colorValue{55.0}  & \colorValue{74.0}  & \colorValue{88.0}  & \colorValue{36.0} & \colorValue{96.0}  & \colorValue{36.0} & \colorValue{0.0} & \colorValue{40.0} & \colorValue{78.0}  &\colorValue{12.0} & \colorValue{55.0}  \\ \cline{2-14}
 & MP3 Compression      & \colorValue{100.0} & \colorValue{75.0}  & \colorValue{80.0}  & \colorValue{88.0}  & \colorValue{80.0} & \colorValue{93.0}  & \colorValue{18.0} & \colorValue{0.0} & \colorValue{71.0} & \colorValue{78.0}  & \colorValue{69.0} & \colorValue{63.0}  \\ \cline{2-14}
 & Pitch Shift          & \colorValue{32.0}  & \colorValue{81.0}  & \colorValue{12.0}  & \colorValue{19.0}  & \colorValue{10.0} & \colorValue{10.0}  & \colorValue{35.0} & \colorValue{0.0} & \colorValue{17.0} & \colorValue{54.0}  & \colorValue{7.0} & \colorValue{9.0}  \\ \cline{2-14}
 & Reverb               & \colorValue{100.0} & \colorValue{99.0}  & \colorValue{100.0} & \colorValue{99.0}  & \colorValue{96.0} & \colorValue{100.0}  & \colorValue{21.0} & \colorValue{0.0} & \colorValue{94.0} & \colorValue{93.0}  & \colorValue{79.0} & \colorValue{91.0}  \\ \cline{2-14}
 & Silence Injection    & \colorValue{100.0} & \colorValue{11.0}  & \colorValue{96.0}  & \colorValue{100.0} & \colorValue{88.0} & \colorValue{100.0}  & \colorValue{22.0} & \colorValue{0.0} & \colorValue{99.0} & \colorValue{94.0}  & \colorValue{81.0} & \colorValue{99.0}  \\ \cline{2-14}
 & Time Stretch         & \colorValue{19.0}  & \colorValue{94.0}  & \colorValue{5.0}   & \colorValue{14.0}  & \colorValue{3.0} & \colorValue{3.0}  & \colorValue{28.0} & \colorValue{0.0} & \colorValue{17.0} & \colorValue{55.0}  & \colorValue{4.0} & \colorValue{2.0}  \\ 
\hline \hline
\multirow{18}{*}{\rotatebox{90}{SPOOF}} 
& \textbf{No Attack}   & \colorValue{93.0}  & \colorValue{85.0}  & \colorValue{98.0}  & \colorValue{93.0}  &\colorValue{95.0} & \colorValue{99.0}  & \colorValue{96.0}  & \colorValue{98.0}  & \colorValue{86.0} & \colorValue{77.0}  &\colorValue{37.0} & \colorValue{87.0} \\ \cline{2-14}
 & Add Background Music & \colorValue{98.0}  & \colorValue{88.0}  & \colorValue{87.0}  & \colorValue{78.0}  & \colorValue{95.0}  & \colorValue{98.0}  & \colorValue{100.0} & \colorValue{84.0}  & \colorValue{87.0} & \colorValue{73.0}  & \colorValue{72.0}  & \colorValue{97.0}  \\ \cline{2-14}
 & Add Background Noise & \colorValue{98.0}  & \colorValue{89.0}  & \colorValue{91.0}  & \colorValue{89.0}  &\colorValue{94.0} & \colorValue{98.0}  & \colorValue{97.0}  & \colorValue{81.0}  & \colorValue{92.0} & \colorValue{80.0}  & \colorValue{57.0}  & \colorValue{95.0}  \\ \cline{2-14}
 & Amplitude Modulation & \colorValue{94.0}  & \colorValue{83.0}  & \colorValue{94.0}  & \colorValue{96.0}  &\colorValue{100.0} & \colorValue{100.0}  & \colorValue{91.0}  & \colorValue{99.0}  & \colorValue{73.0} & \colorValue{58.0}  &\colorValue{94.0} & \colorValue{98.0}  \\ \cline{2-14}
 & Autotune             & \colorValue{95.0}  & \colorValue{88.0}  & \colorValue{96.0}  & \colorValue{93.0}  &\colorValue{100.0} & \colorValue{99.0}  & \colorValue{97.0}  & \colorValue{96.0}  & \colorValue{87.0} & \colorValue{73.0}  &\colorValue{78.0} & \colorValue{93.0}  \\ \cline{2-14}
 & Bit Depth Change     & \colorValue{94.0}  & \colorValue{95.0}  & \colorValue{97.0}  & \colorValue{93.0}  &\colorValue{100.0} & \colorValue{99.0}  & \colorValue{98.0}  & \colorValue{95.0}  & \colorValue{88.0} & \colorValue{67.0}  &\colorValue{66.0} & \colorValue{91.0}  \\ \cline{2-14}
 & Echo                 & \colorValue{98.0}  & \colorValue{91.0}  & \colorValue{97.0}  & \colorValue{95.0}  &\colorValue{100.0} & \colorValue{99.0}  & \colorValue{99.0}  & \colorValue{97.0}  & \colorValue{98.0} & \colorValue{84.0}  &\colorValue{99.0} & \colorValue{100.0}  \\ \cline{2-14}
 & Equalization         & \colorValue{92.0}  & \colorValue{81.0}  & \colorValue{96.0}  & \colorValue{90.0}  &\colorValue{94.0} & \colorValue{99.0}  & \colorValue{93.0}  & \colorValue{94.0}  & \colorValue{87.0} & \colorValue{74.0}  &\colorValue{39.0} & \colorValue{89.0}  \\ \cline{2-14}
 & Freq Minus           & \colorValue{97.0}  & \colorValue{96.0}  & \colorValue{98.0}  & \colorValue{78.0}  &\colorValue{80.0} & \colorValue{100.0}  & \colorValue{96.0}  & \colorValue{99.0}  & \colorValue{95.0} & \colorValue{43.0}  &\colorValue{29.0} & \colorValue{85.0}  \\ \cline{2-14}
 & Freq Plus            & \colorValue{98.0}  & \colorValue{97.0}  & \colorValue{98.0}  & \colorValue{74.0}  &\colorValue{82.0} & \colorValue{98.0}  & \colorValue{95.0}  & \colorValue{99.0}  & \colorValue{95.0} & \colorValue{51.0}  &\colorValue{30.0} & \colorValue{84.0}  \\ \cline{2-14}
 & Gaussian Noise       & \colorValue{100.0} & \colorValue{70.0}  & \colorValue{95.0}  & \colorValue{18.0}  &\colorValue{95.0} & \colorValue{91.0}  & \colorValue{99.0}  & \colorValue{40.0}  & \colorValue{88.0} & \colorValue{6.0}   &\colorValue{93.0} & \colorValue{77.0}  \\ \cline{2-14}
 & High Pass Filter     & \colorValue{37.0}  & \colorValue{20.0}  & \colorValue{90.0}  & \colorValue{18.0}  &\colorValue{91.0} & \colorValue{97.0}  & \colorValue{12.0}  & \colorValue{36.0}  & \colorValue{85.0} & \colorValue{2.0}   &\colorValue{66.0} & \colorValue{92.0}  \\ \cline{2-14}
 & Low Pass Filter      & \colorValue{97.0}  & \colorValue{87.0}  & \colorValue{99.0}  & \colorValue{96.0}  &\colorValue{99.0} & \colorValue{100.0}  & \colorValue{91.0}  & \colorValue{99.0}  & \colorValue{98.0} & \colorValue{71.0}  &\colorValue{95.0} & \colorValue{96.0}  \\ \cline{2-14}
 & MP3 Compression      & \colorValue{97.0}  & \colorValue{90.0}  & \colorValue{98.0}  & \colorValue{97.0}  &\colorValue{88.0} & \colorValue{100.0}  & \colorValue{97.0}  & \colorValue{96.0}  & \colorValue{92.0} & \colorValue{81.0}  &\colorValue{53.0} & \colorValue{98.0}  \\ \cline{2-14}
 & Pitch Shift          & \colorValue{97.0}  & \colorValue{85.0}  & \colorValue{99.0}  & \colorValue{99.0}  &\colorValue{99.0} & \colorValue{100.0}  & \colorValue{92.0}  & \colorValue{91.0}  & \colorValue{99.0} & \colorValue{74.0}  &\colorValue{97.0} & \colorValue{99.0}  \\ \cline{2-14}
 & Reverb               & \colorValue{97.0}  & \colorValue{89.0}  & \colorValue{97.0}  & \colorValue{96.0}  &\colorValue{98.0} & \colorValue{99.0}  & \colorValue{97.0}  & \colorValue{98.0}  & \colorValue{88.0} & \colorValue{73.0}  &\colorValue{52.0} & \colorValue{92.0}  \\ \cline{2-14}
 & Silence Injection    & \colorValue{44.0}  & \colorValue{100.0} & \colorValue{59.0}  & \colorValue{44.0}  &\colorValue{95.0} & \colorValue{74.0}  & \colorValue{80.0}  & \colorValue{100.0} & \colorValue{14.0} & \colorValue{14.0}  &\colorValue{29.0} & \colorValue{17.0}  \\ \cline{2-14}
 & Time Stretch         & \colorValue{98.0}  & \colorValue{85.0}  & \colorValue{100.0} & \colorValue{98.0}  &\colorValue{100.0} & \colorValue{99.0}  & \colorValue{92.0}  & \colorValue{91.0}  & \colorValue{100.0} & \colorValue{72.0}  &\colorValue{99.0} & \colorValue{100.0}  \\ 
\hline
 
\end{tabular} 

    }
    \vspace{0.1cm}
    \caption{The classification accuracy [\%] of the six academic \textbf{open-source models} over the DeePen attacks derived from the ASVspoof 2019 and MLAAD datasets. The consecutive columns refer to Raw PC-DARTS, LCNN, RawGAT-ST, RawNet2, WhisperDF and W2V2 models, respectively. The cell colors are based on the corresponding accuracy: values above $50\%$ are color-coded on a gradient from green ($100\%$ accuracy) to red ($50\%$ accuracy and less). Accuracy below $50\%$ corresponds to a successful attack, as the model performs worse than a random baseline.}
    \label{tab:github}
\end{table*}

\begin{itemize}
    \setlength{\itemindent}{0em}
    \itemsep0em 
    \item \textbf{Dataset difficulty}. The MLAAD dataset is more challenging compared to ASVspoof 2019, as indicated by its much lower average accuracy.
    \item \textbf{Model performance}. Certain models perform better overall. For instance, the W2V model exhibits significantly stronger performance compared to Raw PC-DARTS or WhisperDF.
    \item \textbf{Attack effectiveness}. Some attacks are highly effective, as shown by the ``red'' rows in \Cref{tab:github}. For example, ``Add Background Music,'' ``Add Background Noise,'' and ``Time Stretch'' are particularly successful in disguising bona-fide instances as spoof.
    \item \textbf{Dependence on ground truth}. The efficacy of an attack varies based on the data's label. For example, ``Add Background Music'' is highly effective at generating false positives by disguising bona-fide audio as spoof, but it is less effective at disguising spoof as bona-fide. Conversely, attacks like ``High Pass Filter'' and ``Silence Injection'' are effective at disguising spoofs as bona-fide.
\end{itemize}

In summary, we can see that the available open-source models are very susceptible to some, but not all, of the DeePen attacks.

\begin{table*}[]
    \centering
    \begin{tabular}{|l|l|p{1.1cm}p{1.1cm}p{1.2cm}|p{1.1cm}p{1.1cm}p{1.2cm}|} 
\hline

& &\multicolumn{6}{c|}{\textbf{Accuracy [\%] $\uparrow$}} \\ \cline{3-8}

\multirow{2}{*}{Label} & \multirow{2}{*}{Attack} & \multicolumn{3}{c|}{\textbf{ASVspoof 2019}}       & \multicolumn{3}{c|}{\textbf{MLAAD}}            \\ \cline{3-8} 
                         &                               & System I & System II & System III & System I & System II & System III \\ \hline \hline
\multirow{18}{*}{\rotatebox{90}{BONA-FIDE}}  
 & \textbf{No Attack}   & \colorValue{100.0} & \colorValue{100.0}  & \colorValue{95.0} & \colorValue{100.0} & \colorValue{100.0} & \colorValue{82.0} \\ \cline{2-8} 
 & Add Background Music & \colorValue{98.0}  & \colorValue{98.0}  & \colorValue{77.0} & \colorValue{98.0}  & \colorValue{90.0}  & \colorValue{86.0} \\ \cline{2-8} 
 & Add Background Noise & \colorValue{97.0}  & \colorValue{100.0}  & \colorValue{96.0} & \colorValue{99.0}  & \colorValue{95.0}  & \colorValue{94.0} \\ \cline{2-8}
 & Amplitude Modulation & \colorValue{98.0}  & \colorValue{99.0}  & \colorValue{31.0} & \colorValue{99.0}  & \colorValue{95.0} & \colorValue{12.0} \\ \cline{2-8}
 & Autotune             & \colorValue{29.0}  & \colorValue{99.0}  & \colorValue{51.0} & \colorValue{55.0}  & \colorValue{90.0} & \colorValue{1.0} \\ \cline{2-8} 
 & Bit Depth Change     & \colorValue{100.0}  & \colorValue{98.0} & \colorValue{100.0} & \colorValue{100.0} & \colorValue{84.0} & \colorValue{79.0} \\ \cline{2-8} 
 & Echo                 & \colorValue{53.0}  & \colorValue{93.0}  & \colorValue{22.0}  & \colorValue{42.0}  & \colorValue{96.0} & \colorValue{2.0} \\ \cline{2-8}
 & Equalization         & \colorValue{100.0} & \colorValue{100}  & \colorValue{90.0} & \colorValue{100.0} & \colorValue{99.0} & \colorValue{75.0} \\ \cline{2-8}
 & Freq Minus           & \colorValue{97.0}  & \colorValue{99.0}  & \colorValue{89.0} & \colorValue{98.0}  & \colorValue{87.0} & \colorValue{88.0} \\ \cline{2-8} 
 & Freq Plus            & \colorValue{98.0}  & \colorValue{98.0}  & \colorValue{94.0} & \colorValue{99.0}  & \colorValue{86.0} & \colorValue{77.0} \\ \cline{2-8} 
 & Gaussian Noise       & \colorValue{98.0}  & \colorValue{99.0}  & \colorValue{88.0} & \colorValue{63.0}  & \colorValue{51.0} & \colorValue{28.0} \\ \cline{2-8} 
 & High Pass Filter     & \colorValue{100.0}  & \colorValue{100.0}  & \colorValue{99.0}  & \colorValue{97.0}  & \colorValue{91.0} & \colorValue{60.0} \\ \cline{2-8} 
 & Low Pass Filter      & \colorValue{99.0}  & \colorValue{98.0}  & \colorValue{54.0} & \colorValue{99.0}  & \colorValue{97.0} & \colorValue{55.0} \\ \cline{2-8} 
 & MP3 Compression      & \colorValue{100.0}  & \colorValue{100.0}  & \colorValue{51.0} & \colorValue{97.0}  & \colorValue{87.0} & \colorValue{44.0} \\ \cline{2-8} 
 & Pitch Shift          & \colorValue{16.0}  & \colorValue{32.0}  & \colorValue{10.0} & \colorValue{35.0}  & \colorValue{13.0} & \colorValue{8.0} \\ \cline{2-8}
 & Reverb               & \colorValue{100.0} & \colorValue{100.0}  & \colorValue{93.0} & \colorValue{100.0} & \colorValue{99.0} & \colorValue{72.0} \\ \cline{2-8} 
 & Silence Injection    & \colorValue{100.0} & \colorValue{100.0}  & \colorValue{83.0} & \colorValue{100.0} & \colorValue{99.5} & \colorValue{77.0} \\ \cline{2-8} 
 & Time Stretch         & \colorValue{14.0}  & \colorValue{25.0}  & \colorValue{3.0} & \colorValue{45.0}  & \colorValue{6.0} & \colorValue{2.0} \\  \hline \hline
 
\multirow{18}{*}{\rotatebox{90}{SPOOF}} 
 & \textbf{No Attack}   & \colorValue{99.0}  & \colorValue{96.0}  & \colorValue{98.0} & \colorValue{94.0}  & \colorValue{99.0}  & \colorValue{100.0} \\ \cline{2-8}
 & Add Background Music & \colorValue{78.50}  & \colorValue{80.0}  & \colorValue{78.0} & \colorValue{59.0}  & \colorValue{83.0}  & \colorValue{54.0} \\ \cline{2-8}
 & Add Background Noise & \colorValue{86.0}  & \colorValue{86.0}  & \colorValue{83.0} & \colorValue{79.0}  & \colorValue{88.0}  & \colorValue{59.0} \\ \cline{2-8}
 & Amplitude Modulation & \colorValue{99.0}  & \colorValue{97.0}  & \colorValue{100.0} & \colorValue{88.0}  & \colorValue{99.0}  & \colorValue{100.0} \\ \cline{2-8} 
 & Autotune             & \colorValue{100.0} & \colorValue{100.0}  & \colorValue{99.0} & \colorValue{100.0} & \colorValue{100.0}  & \colorValue{100.0} \\ \cline{2-8} 
 & Bit Depth Change     & \colorValue{100.0} & \colorValue{99.0}  & \colorValue{98.0} & \colorValue{88.0}  & \colorValue{100.0} & \colorValue{90.0} \\ \cline{2-8} 
 & Echo                 & \colorValue{98.0}  & \colorValue{96.0}  & \colorValue{100.0} & \colorValue{98.0}  & \colorValue{90.0} & \colorValue{99.0} \\ \cline{2-8}
 & Equalization         & \colorValue{99.0}  & \colorValue{93.0}  & \colorValue{96.0} & \colorValue{95.0}  & \colorValue{99.0}  & \colorValue{98.0} \\ \cline{2-8} 
 & Freq Minus           & \colorValue{96.0}  & \colorValue{89.0}  & \colorValue{92.0} & \colorValue{86.0}  & \colorValue{95.0}  & \colorValue{87.0} \\ \cline{2-8} 
 & Freq Plus            & \colorValue{97.0}  & \colorValue{90.0}  & \colorValue{91.0} & \colorValue{87.0}  & \colorValue{96.0}  & \colorValue{86.0} \\ \cline{2-8} 
 & Gaussian Noise       & \colorValue{82.0}  & \colorValue{80.0}  & \colorValue{90.0} & \colorValue{76.0}  & \colorValue{93.0}  & \colorValue{92.0} \\ \cline{2-8}
 & High Pass Filter     & \colorValue{69.0}  & \colorValue{57.0}  & \colorValue{68.0} & \colorValue{13.0}  & \colorValue{81.0}  & \colorValue{48.0} \\ \cline{2-8}
 & Low Pass Filter      & \colorValue{90.0}  & \colorValue{88.0}  & \colorValue{98.0} & \colorValue{78.0}  & \colorValue{98.0}  & \colorValue{95.0} \\ \cline{2-8} 
 & MP3 Compression      & \colorValue{98.0}  & \colorValue{99.0}  & \colorValue{100.0} & \colorValue{93.0}  & \colorValue{98.0}  & \colorValue{100.0} \\ \cline{2-8} 
 & Pitch Shift          & \colorValue{85.0}  & \colorValue{98.0}  & \colorValue{99.0} & \colorValue{77.0}  & \colorValue{99.0}  & \colorValue{100.0} \\ \cline{2-8} 
 & Reverb               & \colorValue{98.0}  & \colorValue{96.0}  & \colorValue{98.0} & \colorValue{94.0}  & \colorValue{99.0}  & \colorValue{100.0} \\ \cline{2-8} 
 & Silence Injection    & \colorValue{98.0}  & \colorValue{98.0}  & \colorValue{100.0} & \colorValue{87.0}  & \colorValue{99.0}  & \colorValue{99.0} \\ \cline{2-8} 
 & Time Stretch         & \colorValue{84.0}  & \colorValue{100.0}  & \colorValue{100.0} & \colorValue{69.0}  & \colorValue{100.0}  & \colorValue{100.0} \\  \hline
\end{tabular}

    \vspace{0.1cm}
    \caption{The classification accuracy [\%] of the tree \textbf{commercial deepfake detection systems} over the DeePen penetration test attacks derived from the ASVspoof 2019 and MLAAD datasets. The cell colors are based on the corresponding accuracy: values above $50\%$ are color-coded on a gradient from green ($100\%$ accuracy) to red ($50\%$ accuracy and less). Accuracy below $50\%$ corresponds to a successful attack, as the model performs worse than a random baseline.}
    \label{tab:commercial}
\end{table*}

\subsection{Commercial deepfake detection models}
\label{ss:commercial-models}

Secondly, we evaluate commercial, proprietary models accessed through a Software-as-a-Service (SaaS) framework. These solutions are provided by some of the established vendors in the market, available via API, and are widely used by individuals, journalists, businesses, and government organizations. Due to confidentiality agreements, we refrain from disclosing the vendor identities.

The classification results of the commercial audio deepfake detection systems in the DeePen penetration test are shown in \Cref{tab:commercial}. 
We can infer the following.
\begin{itemize}
    \item First, we note that some models perform quite well, while others show overall poor performance. For example, System I and System II are vulnerable to only a few attacks, while System III is considerably more exposed. System IV performs poorly across the board and, as indicated by the low scores in the row ``No Attack'', exhibits a strong bias towards predicting ``spoof'' regardless of input, even in the absence of an attack.
    \item Second, while System I and System II are fairly resilient, several attacks still reliably deceive these models. For example, attacks with ``Pitch Shift'' and ``Time Stretch'' can degrade the average accuracy by $16$ to $32$ percentage points, far worse than random guessing. Similarly, the ``High Pass Filter'' attack can be used to disguise spoofed files as authentic. Less effective, but still concerning, is the ``Add Background Music'' attack, which, while not fully incapacitating the detection model, significantly reduces performance (e.g., from $99$ percent accuracy to approximately $78$ percent for System I).
\end{itemize}

Thus, while some SaaS solutions demonstrate strong performance and resilience against DeePen attacks, there is always at least one attack capable of deceiving the system.

\subsection{Adaptive Defense}
\label{ss:adaptive}

Given the prior results, we perform the ``Adaptive Defense'' evaluation step. Here, we evaluate DeePen penetration test against an adaptive defender who is aware of the attacks and incorporates this knowledge to improve their detection model. 
In this context, the attacks of DeePen can be used as a ``defense'' against the aforementioned ``attacks'' by re-training the defender's model on part or the complete set of DeePen's audio attacks.
In this evaluation step, we anticipate improved robustness against attacks, and are interested in whether adaptive retraining is sufficient to completely neutralize the DeePen attacks.

\begin{table*}[]
    \centering
    \resizebox{.99\textwidth}{!}{
    \begin{tabular}{llllllllllllllllll|ll}
attack $\rightarrow$ & \rotatebox{90}{Add Background Music} & \rotatebox{90}{Add Background Noise} & \rotatebox{90}{Amplitude Modulation} & \rotatebox{90}{Autotune} & \rotatebox{90}{Bit Depth Change} & \rotatebox{90}{Echo} & \rotatebox{90}{Equalization} & \rotatebox{90}{Freq Minus} & \rotatebox{90}{Freq Plus} & \rotatebox{90}{Gaussian Noise} & \rotatebox{90}{High Pass Filter} & \rotatebox{90}{Low Pass Filter} & \rotatebox{90}{MP3 Compression} & \rotatebox{90}{Pitch Shift} & \rotatebox{90}{Reverb} & \rotatebox{90}{Silence Injection} & \rotatebox{90}{Time Stretch} & \rotatebox{90}{No Attack} & \rotatebox{90}{\textbf{Mean}} \\
adaptive defense $\downarrow$ &  &  &  &  &  &  &  &  &  &  &  &  &  &  &  &  &  &  &  \\
\midrule
Add Background Music & \deltaValue{+7.8} & \deltaValue{+7.2} & \deltaValue{+1.2} & \deltaValue{-4.9} & \deltaValue{+1.0} & \deltaValue{+4.7} & \deltaValue{-0.4} & \deltaValue{-0.3} & \deltaValue{+1.7} & \deltaValue{+7.5} & \deltaValue{+4.1} & \deltaValue{+1.0} & \deltaValue{+0.0} & \deltaValue{+0.0} & \deltaValue{+0.4} & \deltaValue{+0.9} & \deltaValue{-0.1} & \deltaValue{+0.2} & \deltaValue{+1.8} \\
Add Background Noise & \deltaValue{+7.5} & \deltaValue{+8.2} & \deltaValue{+3.6} & \deltaValue{-0.4} & \deltaValue{+1.3} & \deltaValue{+3.3} & \deltaValue{+0.2} & \deltaValue{+2.7} & \deltaValue{+3.0} & \deltaValue{+15.5} & \deltaValue{+6.5} & \deltaValue{+1.0} & \deltaValue{+1.8} & \deltaValue{+0.0} & \deltaValue{+0.0} & \deltaValue{+0.4} & \deltaValue{+0.0} & \deltaValue{+0.2} & \deltaValue{+3.0} \\
Amplitude Modulation & \deltaValue{+1.3} & \deltaValue{+4.8} & \deltaValue{+14.7} & \deltaValue{-3.0} & \deltaValue{+0.1} & \deltaValue{+1.2} & \deltaValue{-0.3} & \deltaValue{-1.0} & \deltaValue{-1.3} & \deltaValue{+8.1} & \deltaValue{+3.4} & \deltaValue{+0.8} & \deltaValue{-1.4} & \deltaValue{+0.2} & \deltaValue{+0.2} & \deltaValue{+0.6} & \deltaValue{+0.0} & \deltaValue{+0.1} & \deltaValue{+1.6} \\
Autotune & \deltaValue{-1.1} & \deltaValue{-0.7} & \deltaValue{+3.1} & \deltaValue{+26.9} & \deltaValue{+0.2} & \deltaValue{+1.1} & \deltaValue{-0.1} & \deltaValue{-2.0} & \deltaValue{-1.3} & \deltaValue{+0.0} & \deltaValue{+3.6} & \deltaValue{+0.1} & \deltaValue{-1.3} & \deltaValue{+0.0} & \deltaValue{-0.6} & \deltaValue{+0.3} & \deltaValue{+0.0} & \deltaValue{-0.1} & \deltaValue{+1.6} \\
Bit Depth Change & \deltaValue{+2.2} & \deltaValue{+3.0} & \deltaValue{-2.4} & \deltaValue{-1.0} & \deltaValue{+1.5} & \deltaValue{+4.0} & \deltaValue{-0.4} & \deltaValue{-3.5} & \deltaValue{-1.6} & \deltaValue{+7.5} & \deltaValue{+6.3} & \deltaValue{+0.2} & \deltaValue{-0.9} & \deltaValue{+0.0} & \deltaValue{-0.2} & \deltaValue{+0.0} & \deltaValue{+0.0} & \deltaValue{-0.1} & \deltaValue{+0.8} \\
Echo & \deltaValue{-1.2} & \deltaValue{+2.0} & \deltaValue{+1.4} & \deltaValue{+1.9} & \deltaValue{+0.7} & \deltaValue{+32.6} & \deltaValue{+0.0} & \deltaValue{-1.6} & \deltaValue{+0.1} & \deltaValue{+5.0} & \deltaValue{+4.8} & \deltaValue{-0.2} & \deltaValue{+1.1} & \deltaValue{+0.1} & \deltaValue{-0.4} & \deltaValue{+0.6} & \deltaValue{-0.1} & \deltaValue{+0.1} & \deltaValue{+2.6} \\
Equalization & \deltaValue{+2.3} & \deltaValue{+1.8} & \deltaValue{+1.3} & \deltaValue{-1.5} & \deltaValue{+0.1} & \deltaValue{-1.1} & \deltaValue{+0.0} & \deltaValue{-0.3} & \deltaValue{+0.6} & \deltaValue{+8.9} & \deltaValue{+6.5} & \deltaValue{+1.9} & \deltaValue{-1.9} & \deltaValue{+0.1} & \deltaValue{+0.4} & \deltaValue{+0.6} & \deltaValue{+0.0} & \deltaValue{+0.3} & \deltaValue{+1.1} \\
Freq Minus & \deltaValue{+1.9} & \deltaValue{+3.8} & \deltaValue{+4.2} & \deltaValue{+1.9} & \deltaValue{+0.7} & \deltaValue{+1.8} & \deltaValue{+0.0} & \deltaValue{+3.2} & \deltaValue{+4.4} & \deltaValue{+12.7} & \deltaValue{+3.9} & \deltaValue{+0.5} & \deltaValue{+4.0} & \deltaValue{+0.0} & \deltaValue{+0.2} & \deltaValue{+0.4} & \deltaValue{+0.0} & \deltaValue{+0.1} & \deltaValue{+2.4} \\
Freq Plus & \deltaValue{+3.7} & \deltaValue{+6.0} & \deltaValue{+3.6} & \deltaValue{-1.8} & \deltaValue{+1.5} & \deltaValue{-1.0} & \deltaValue{+0.3} & \deltaValue{+3.1} & \deltaValue{+3.6} & \deltaValue{+18.3} & \deltaValue{+2.2} & \deltaValue{+1.3} & \deltaValue{-0.7} & \deltaValue{+0.1} & \deltaValue{+0.6} & \deltaValue{+0.8} & \deltaValue{+0.0} & \deltaValue{+0.4} & \deltaValue{+2.3} \\
Gaussian Noise & \deltaValue{+4.7} & \deltaValue{+7.8} & \deltaValue{+0.7} & \deltaValue{-2.7} & \deltaValue{+1.0} & \deltaValue{+0.1} & \deltaValue{+0.2} & \deltaValue{+1.3} & \deltaValue{+3.6} & \deltaValue{+21.6} & \deltaValue{+8.8} & \deltaValue{+1.7} & \deltaValue{-1.9} & \deltaValue{+0.1} & \deltaValue{+0.5} & \deltaValue{+1.0} & \deltaValue{+0.0} & \deltaValue{+0.0} & \deltaValue{+2.7} \\
High Pass Filter & \deltaValue{-0.1} & \deltaValue{+1.9} & \deltaValue{+5.8} & \deltaValue{+3.4} & \deltaValue{+1.3} & \deltaValue{+0.0} & \deltaValue{-0.3} & \deltaValue{+0.6} & \deltaValue{+2.6} & \deltaValue{+6.0} & \deltaValue{+36.3} & \deltaValue{+0.4} & \deltaValue{+4.2} & \deltaValue{+0.0} & \deltaValue{-0.3} & \deltaValue{+0.5} & \deltaValue{+0.0} & \deltaValue{+0.0} & \deltaValue{+3.5} \\
Low Pass Filter & \deltaValue{+3.5} & \deltaValue{+4.9} & \deltaValue{-0.3} & \deltaValue{-1.5} & \deltaValue{+0.9} & \deltaValue{-2.1} & \deltaValue{-0.2} & \deltaValue{+0.0} & \deltaValue{+1.6} & \deltaValue{+11.4} & \deltaValue{+3.4} & \deltaValue{+4.8} & \deltaValue{-1.8} & \deltaValue{+0.0} & \deltaValue{+0.4} & \deltaValue{+0.6} & \deltaValue{+0.0} & \deltaValue{+0.2} & \deltaValue{+1.4} \\
MP3 Compression & \deltaValue{-1.8} & \deltaValue{+1.0} & \deltaValue{+1.0} & \deltaValue{+3.1} & \deltaValue{+0.9} & \deltaValue{+1.7} & \deltaValue{-0.5} & \deltaValue{+0.5} & \deltaValue{+0.6} & \deltaValue{+9.8} & \deltaValue{+1.2} & \deltaValue{-0.9} & \deltaValue{+7.6} & \deltaValue{+0.0} & \deltaValue{-0.2} & \deltaValue{+0.2} & \deltaValue{+0.0} & \deltaValue{+0.0} & \deltaValue{+1.3} \\
Pitch Shift & \deltaValue{+3.1} & \deltaValue{+4.0} & \deltaValue{+2.4} & \deltaValue{+0.4} & \deltaValue{+0.5} & \deltaValue{+4.8} & \deltaValue{-0.3} & \deltaValue{-1.7} & \deltaValue{-0.1} & \deltaValue{+2.1} & \deltaValue{+5.4} & \deltaValue{+0.5} & \deltaValue{+5.2} & \deltaValue{+23.1} & \deltaValue{+0.4} & \deltaValue{+0.6} & \deltaValue{+29.1} & \deltaValue{+0.0} & \deltaValue{+4.4} \\
Reverb & \deltaValue{+3.2} & \deltaValue{+3.6} & \deltaValue{+2.8} & \deltaValue{+1.3} & \deltaValue{-0.2} & \deltaValue{+2.5} & \deltaValue{-1.3} & \deltaValue{-2.4} & \deltaValue{-1.1} & \deltaValue{+6.7} & \deltaValue{+3.8} & \deltaValue{-0.7} & \deltaValue{-0.5} & \deltaValue{+0.1} & \deltaValue{-0.5} & \deltaValue{-1.2} & \deltaValue{+0.0} & \deltaValue{-1.1} & \deltaValue{+0.8} \\
Silence Injection & \deltaValue{-1.0} & \deltaValue{+0.4} & \deltaValue{-0.4} & \deltaValue{+1.9} & \deltaValue{-0.9} & \deltaValue{-1.2} & \deltaValue{-0.3} & \deltaValue{-1.9} & \deltaValue{+0.0} & \deltaValue{+2.2} & \deltaValue{+5.5} & \deltaValue{+0.6} & \deltaValue{-1.1} & \deltaValue{+0.1} & \deltaValue{+0.3} & \deltaValue{+0.3} & \deltaValue{+0.0} & \deltaValue{-0.3} & \deltaValue{+0.2} \\
Time Stretch & \deltaValue{+0.2} & \deltaValue{+2.4} & \deltaValue{+2.8} & \deltaValue{-0.7} & \deltaValue{-1.1} & \deltaValue{+6.0} & \deltaValue{-0.2} & \deltaValue{-2.0} & \deltaValue{-0.5} & \deltaValue{+2.1} & \deltaValue{+5.8} & \deltaValue{+0.8} & \deltaValue{+3.8} & \deltaValue{+24.1} & \deltaValue{+0.1} & \deltaValue{+1.0} & \deltaValue{+33.7} & \deltaValue{-0.3} & \deltaValue{+4.3} \\ \hline
\textbf{No Defense} & 84.1 & 87.3 & 82.2 & 66.7 & 98.2 & 59.9 & 99.3 & 95.2 & 94.8 & 74.7 & 50.8 & 89.8 & 90.1 & 55.0 & 99.2 & 98.3 & 51.5 & 99.4 & 82.0 \\
 \textbf{Retrain All}& \deltaValue{+9.2} & \deltaValue{+9.6} & \deltaValue{+15.1} & \deltaValue{+28.9} & \deltaValue{+1.7} & \deltaValue{+33.7} & \deltaValue{-0.2} & \deltaValue{+3.9} & \deltaValue{+4.5} & \deltaValue{+21.4} & \deltaValue{+34.9} & \deltaValue{+3.4} & \deltaValue{+8.2} & \deltaValue{+25.2} & \deltaValue{-0.8} & \deltaValue{+0.5} & \deltaValue{+33.0} & \deltaValue{-0.3} & \deltaValue{+12.9} \\
 \textbf{Retrain Selected}& \deltaValue{+10.4} & \deltaValue{+8.9} & \deltaValue{+15.3} & \deltaValue{+27.8} & \deltaValue{+1.7} & \deltaValue{+33.1} & \deltaValue{+0.2} & \deltaValue{+2.7} & \deltaValue{+4.0} & \deltaValue{+22.2} & \deltaValue{+36.3} & \deltaValue{+0.0} & \deltaValue{+8.7} & \deltaValue{+23.9} & \deltaValue{-0.4} & \deltaValue{+1.1} & \deltaValue{+31.5} & \deltaValue{-0.3} & \deltaValue{+12.6} \\

\bottomrule
\end{tabular}

    }
    \vspace{0.1cm}
    \caption{Performance of an adaptive defender over the ASVspoof 2019 dataset. The defender system is retrained  on the complete DeePen dataset (`Retrain All'), or a subset of the attacks. Performance of the baseline system, without retraining on DeePen, is shown as `No Defense'. Results are averaged over five independent trials, and denote the average accuracy modification relative to the `No Defense' scenario.}
    \label{tab:asv19_adaptive_delta}
\end{table*}

\subsubsection{Training Setup}
To evaluate the efficacy of an adaptive defense scheme, we must train, attack, and defend an audio deepfake detection model ourselves. For this purpose, we selected the W2V2 deepfake detection model from \Cref{ss:open-models}. W2V2 is widely used in the scientific community, it is open-source, and has demonstrated strong performance in our prior evaluation step (see \Cref{tab:github}).

We train this model on the train partitions of the ASVspoof 2019 and MLAAD datasets using various configurations: (i) without adaptive defense; (ii) with adaptive defense using just one of the $K=17$ proposed attacks; (iii) with all attacks simultaneously; (iv) or with a subset of attacks.
The defense model is trained for $100$ epochs with early stopping on the training loss. We use a learning rate of $4 \times 10^{-5}$, a batch size of $16$, and the Adam optimizer~\cite{kingma2015}. Reported values are computed over five independent, identical trials and presented as the average accuracy of the deepfake system's ability to classify real vs fake samples.

\subsubsection{Results}The results of the adaptive defense over the test sets of the two datasets are shown in \Cref{tab:asv19_adaptive_delta}. For each DeePen defense (row) and each attack (column), the table shows the relative change in performance compared to the baseline ``No Defense''). For example, in the top row, the first two values of $+7.8$ and $+7.2$ indicate that using the ``Add Background Music'' defense improves the model's performance by $+7.8$ and $+7.2$ percent in absolute accuracy against the ``Add Background Music'' and ``Add Background Noise'' attacks, respectively. This demonstrates that a single defense can be effective not only against the corresponding attack but also against other, seemingly unrelated attacks.

\subsubsection{Choice of Adaptive Defenses}

One challenge that an adaptive defender may face is that of deciding which of the $17$ possible DeePen modifications to use as a defense, i.e. for adaptive retraining. 
The naive approach, commonly adopted in the literature~\cite{rawboost}, is to retrain the model using all available modifications, that is, to use all available defenses. 
However, it may be beneficial to select only a minimal set of defensive strategies. 
This reduces data augmentation and training time, and also helps to understand what the deepfake detection model may learn. 
We employ a greedy approach, as described in \Cref{greed_threshold}. 
Intuitively, for $n = m = 17$ potential attacks and defenses, our objective is to select fewer than $m$ defenses that perform better or at least equal compared to using all $m$ defenses.

To find this minimal set of defenses, we first evaluate the performance of the adaptive defender's model trained on each defense individually, resulting in a performance matrix $A \in \mathbb{R}^{m \times n}$. 
Here, the entry at position $i, j$ indicated the performance of a model, employing defense $i$, against attack $j$.
See \Cref{tab:asv19_adaptive_delta} for an example: rows correspond to defenses, and columns correspond to the evaluated attacks.

The optimal subset of defenses $S$ is then selected as follows:
For a defense $i$ to be included in the adaptive defender's optimal set $S$, we require two conditions:
(i) first, it should defend better than any other defense against at least one attack; (ii)
second, the defense should also outperform the baseline ``no defense'' by $L$ percent.
Put differently, we require that the corresponding defense provides unique coverage against at least one attack such that (i) the defense in question cannot be substituted by another defense (i.e., it is maximally effective); and (ii) the defense in question is at least somewhat effective (i.e. exceeds baseline ``No Defense'' by at least $L$ percent).
We chose $L = 5\%$, i.e. we require that a defense improves the model's performance by at least $5\%$.

\begin{algorithm}[t]
\caption{Greedy Algorithm for Adaptive Defense Selection}
\begin{algorithmic}
\STATE \textbf{Input:} Matrix $A \in \mathbb{R}^{m \times n}$, where $m$ is the number of possible adaptive augmentations and $n$ is the number of attacks, constant $L$
\STATE \textbf{Output:} Subset of selected rows $S$
\STATE $S \gets \emptyset$ 

\FOR{$i = 1$ \TO $m$}
    \FOR{$j = 1$ \TO $n$}
        \IF{$A[i,j] \geq \max(A[:,j])$ \AND $A[i,j] \geq L$}
            \STATE $S \gets S \cup \{i\}$ 
            \STATE \textbf{break} 
        \ENDIF
    \ENDFOR
\ENDFOR
\RETURN $S$
\end{algorithmic}
\label{greed_threshold}
\end{algorithm}

The results of this approach are shown in \Cref{tab:asv19_adaptive_delta}.
The performance of the model when all defenses are applied is shown in the row labeled ``Retrain All''. The performance for individual defenses appear to aggregate, resulting in a model that is more robust against a variety of attacks. For example, compared to the baseline, retraining with all defenses improves overall accuracy by an average of $+12.9\%$, with gains as high as $+34.9\%$ for the ``High Pass Filter'' attack.
Note that the average accuracy for a ``Retrain All'' defended model is $94.9\%$ (see rightmost column in \Cref{tab:asv19_adaptive_delta}).
This is approximately $5\%$ less than the model performance in the absence of an attack (where the average accuracy is $99.4\%$). Thus, while the ``Retrain All'' defense is undoubtedly beneficial and yields a $+12.9\%$ increase in average accuracy compared to the baseline ``No Defense'', the impact of the DeePen attacks cannot be entirely mitigated.

\textbf{Comparison Between W2V2 Checkpoints.}
The publicly available W2V2 checkpoint demonstrated superior performance on the ASVspoof 2019 dataset (as reported in \Cref{tab:github}) compared to the W2V2 model presented in \Cref{tab:asv19_adaptive_delta}. The primary difference between these systems lies in their training process. The model in \Cref{tab:asv19_adaptive_delta} was trained on both the ASVspoof 2019 and MLAAD datasets, whereas the model in \Cref{tab:github} was trained exclusively on ASVspoof 2019. However, the latter incorporated the RawBoost data augmentation technique, which has been shown to improve performance~\cite{rawboost}. For this reason, we exclude it.

\subsubsection{Understanding the Defense}
As outlined previously, our objective is to obtain a deeper understanding of the impact of various defenses. 
We start by examining the results in \Cref{tab:asv19_adaptive_delta}, where we observe that certain attacks appear to be able to substitute each other. 
For instance, both ``Time Stretch'' and ``Pitch Shift'' effectively counteract each other. 
This is reasonable given that a time stretch naturally induces a pitch alteration in the audio file.
However, given that a ``Pitch Shift'' does not induce a change in play speed, we can assume that the audio deepfake detection model mostly focuses on the change in pitch, rather than playback speed. 
Thus, the ``Time Stretch'' defense protects against ``Pitch Shift'' attacks, and vice versa.
Furthermore, the ``Gaussian Noise'' attack can be mitigated by multiple defenses (as indicated by the blue highlights in the ``Gaussian Noise'' column).
Several other surprising pairs can be found (highlighted in blue). For example, the ``Gaussian Noise'' defense is only marginally weaker than the ``Add Background Noise'' defense in defending against ``Add Background Noise'' ($+7.8\%$ vs $+8.2\%$ absolute accuracy improvement).

The observation that not every attack requires a mirrored defense prompts us to investigate whether a minimal set of defenses could suffice to counter all attacks.
Applying the greedy algorithm from \Cref{greed_threshold}, we can identify a smaller and more representative set of defenses that performs just as well as the naive approach of using all defenses. 
\Cref{tab:selected_asv} lists the nine defenses that can replace the remaining eight that were not selected. Using this subset of defenses, we achieve a mean and maximum performance comparable to the naive ``Retrain All'' approach (see the last row, ``Retrain Selected'', in \Cref{tab:asv19_adaptive_delta}).
For example, even though we do not use the defense ``Frequency Plus'' in the ``Retrain Selected'' scenario, the model performance is comparable to the ``Retrain all'' scenario (plus $4.0\%$ vs. $4.5$ accuracy).
Similarly for the ``Pitch Shift'' or ``Bit Depth Change'' attack.
In summary, while some attacks require exactly the same defense to be mitigated, not all do.
These findings are corroborated by similar experiments on MLAAD, as shown in \Cref{tab:mlaad_adaptive_delta} in the appendix. 
For a more detailed analysis of how defenses affect the performance of the model across specific labels, refer to \Cref{tab:asv19_adaptive_carrier_delta}.

\begin{table}[]
    \centering
    \begin{tabular}{|l|}
\hline
\textbf{Audio Attacks} \\ \hline
Add Background Noise \\ 
Add Background Music \\ 
Amplitude Modulation \\ 
Autotune \\ 
Echo \\ 
Gaussian Noise \\ 
High Pass Filter \\ 
MP3 Compression \\ 
Time Stretch \\ \hline
\end{tabular}


    \vspace{0.1cm}
    \caption{Subset of audio attacks as identified by \Cref{greed_threshold} for the results of DeePen attacks on ASVspoof 2019, as shown in \Cref{tab:asv19_adaptive_delta}.}
    \label{tab:selected_asv}
\end{table}

\section{Discussions and conclusion}

\subsection{Perceptibility of Attacks}

To evaluate the perceptibility of the attacks, we invite the reader to listen to sample audio files\footnote{\url{https://github.com/Fraunhofer-AISEC/DeePen}}. It is important to note that imperceptibility, although desirable, is not a necessary condition for an attack to be effective.
First, some attacks (such as background noise or music), though highly perceptible, do not diminish the credibility of the audio file.
Second, certain attacks that degrade audio quality (such as Gaussian noise or bandpass filtering) can be used~\cite{ste} to enhance the plausibility of the deepfake audio. For example, they may support the narrative that the recording was captured by a hidden microphone or intercepted during a telephone conversation.
Third, even clearly distorted audio can be exploited to overwhelm response and inspection teams with false positives, thereby eroding trust in the system. This is analogous to an antivirus program that, when producing too many false positives, becomes ineffective.


\subsection{Conclusion}
Our research highlights the vulnerabilities in current state-of-the-art deepfake detection systems, demonstrating that simple attacks can reliably bypass these systems. 
Through the introduction of DeePen, we provide a framework for evaluating the robustness of commercial and open-source models in a penetration test setup.
Our findings emphasize the importance of adaptive retraining to mitigate some attacks, but also show that certain attacks still remain effective.
Additionally, our identification of a minimal representative set of adaptive augmentations presents a more efficient approach for defending against deepfake attacks. 
The need for continued innovation in this space is clear, as robust and adaptive detection systems are critical to ensuring the integrity of audio-based communications.

\bibliographystyle{IEEEtran}
\bibliography{bib}

\appendices

\begin{table*}[]
    \centering
    \resizebox{.99\textwidth}{!}{
    \begin{tabular}{llllllllllllllllll|ll}
\toprule
attack $\rightarrow$ & \rotatebox{90}{Add Background Music} & \rotatebox{90}{Add Background Noise} & \rotatebox{90}{Amplitude Modulation} & \rotatebox{90}{Autotune} & \rotatebox{90}{Bit Depth Change} & \rotatebox{90}{Echo} & \rotatebox{90}{Equalization} & \rotatebox{90}{Freq Minus} & \rotatebox{90}{Freq Plus} & \rotatebox{90}{Gaussian Noise} & \rotatebox{90}{High Pass Filter} & \rotatebox{90}{Low Pass Filter} & \rotatebox{90}{MP3 Compression} & \rotatebox{90}{Pitch Shift} & \rotatebox{90}{Reverb} & \rotatebox{90}{Silence Injection} & \rotatebox{90}{Time Stretch} & \rotatebox{90}{No Attack} & \rotatebox{90}{\textbf{Mean}} \\
adaptive defense $\downarrow$ &  &  &  &  &  &  &  &  &  &  &  &  &  &  &  &  &  &  &  \\
\midrule
Add Background Music & \deltaValue{+12.0} & \deltaValue{+3.6} & \deltaValue{+1.3} & \deltaValue{-12.8} & \deltaValue{+1.2} & \deltaValue{+7.2} & \deltaValue{-0.3} & \deltaValue{-0.7} & \deltaValue{-2.7} & \deltaValue{+3.5} & \deltaValue{+1.8} & \deltaValue{+2.2} & \deltaValue{-0.4} & \deltaValue{+1.4} & \deltaValue{+0.6} & \deltaValue{+0.3} & \deltaValue{+1.0} & \deltaValue{+0.4} & \deltaValue{+1.1} \\
Add Background Noise & \deltaValue{+7.5} & \deltaValue{+4.1} & \deltaValue{-0.9} & \deltaValue{-7.8} & \deltaValue{-0.4} & \deltaValue{+2.9} & \deltaValue{-1.0} & \deltaValue{+9.1} & \deltaValue{+8.2} & \deltaValue{+13.0} & \deltaValue{+2.2} & \deltaValue{+1.8} & \deltaValue{-0.6} & \deltaValue{+0.0} & \deltaValue{-0.1} & \deltaValue{-0.6} & \deltaValue{-0.2} & \deltaValue{-0.1} & \deltaValue{+2.1} \\
Amplitude Modulation & \deltaValue{-2.3} & \deltaValue{-0.2} & \deltaValue{+16.6} & \deltaValue{-11.7} & \deltaValue{-0.5} & \deltaValue{+3.9} & \deltaValue{-1.1} & \deltaValue{-1.7} & \deltaValue{-1.1} & \deltaValue{+4.0} & \deltaValue{+1.4} & \deltaValue{+1.9} & \deltaValue{-1.0} & \deltaValue{+0.0} & \deltaValue{+0.4} & \deltaValue{-0.1} & \deltaValue{-0.5} & \deltaValue{-0.3} & \deltaValue{+0.4} \\
Autotune & \deltaValue{-3.7} & \deltaValue{-3.1} & \deltaValue{-3.4} & \deltaValue{+15.9} & \deltaValue{-0.1} & \deltaValue{+1.0} & \deltaValue{-1.3} & \deltaValue{-2.1} & \deltaValue{-4.3} & \deltaValue{+2.0} & \deltaValue{+1.3} & \deltaValue{+2.4} & \deltaValue{-1.7} & \deltaValue{-0.2} & \deltaValue{+0.1} & \deltaValue{+0.0} & \deltaValue{-0.1} & \deltaValue{+0.1} & \deltaValue{+0.2} \\
Bit Depth Change & \deltaValue{-0.7} & \deltaValue{-0.2} & \deltaValue{+0.0} & \deltaValue{-6.4} & \deltaValue{+0.0} & \deltaValue{+4.0} & \deltaValue{-2.3} & \deltaValue{-0.1} & \deltaValue{-0.2} & \deltaValue{+4.3} & \deltaValue{+0.5} & \deltaValue{+0.4} & \deltaValue{-0.1} & \deltaValue{-0.2} & \deltaValue{-0.2} & \deltaValue{-0.6} & \deltaValue{-0.5} & \deltaValue{-0.5} & \deltaValue{-0.2} \\
Echo & \deltaValue{+0.6} & \deltaValue{-2.4} & \deltaValue{+2.0} & \deltaValue{+1.4} & \deltaValue{-2.3} & \deltaValue{+38.1} & \deltaValue{-1.6} & \deltaValue{+0.5} & \deltaValue{-1.6} & \deltaValue{+2.8} & \deltaValue{+3.3} & \deltaValue{-0.5} & \deltaValue{-0.5} & \deltaValue{+0.2} & \deltaValue{-0.6} & \deltaValue{-0.4} & \deltaValue{+1.5} & \deltaValue{-1.1} & \deltaValue{+2.2} \\
Equalization & \deltaValue{+1.3} & \deltaValue{-2.0} & \deltaValue{-3.5} & \deltaValue{-14.1} & \deltaValue{+0.1} & \deltaValue{-2.3} & \deltaValue{-0.4} & \deltaValue{-3.7} & \deltaValue{-4.9} & \deltaValue{+4.0} & \deltaValue{+2.6} & \deltaValue{+3.5} & \deltaValue{-1.6} & \deltaValue{+0.1} & \deltaValue{+0.8} & \deltaValue{+0.2} & \deltaValue{-1.0} & \deltaValue{+0.3} & \deltaValue{-1.1} \\
Freq Minus & \deltaValue{+2.3} & \deltaValue{+1.7} & \deltaValue{-2.8} & \deltaValue{-7.3} & \deltaValue{+0.5} & \deltaValue{+0.6} & \deltaValue{-0.2} & \deltaValue{+11.6} & \deltaValue{+13.6} & \deltaValue{+5.4} & \deltaValue{+0.0} & \deltaValue{+2.9} & \deltaValue{+2.0} & \deltaValue{-0.1} & \deltaValue{+0.6} & \deltaValue{-0.4} & \deltaValue{-0.3} & \deltaValue{+0.2} & \deltaValue{+1.7} \\
Freq Plus & \deltaValue{+5.0} & \deltaValue{+0.3} & \deltaValue{-1.6} & \deltaValue{-9.7} & \deltaValue{+0.0} & \deltaValue{-1.9} & \deltaValue{-0.6} & \deltaValue{+11.9} & \deltaValue{+12.3} & \deltaValue{+8.0} & \deltaValue{+1.0} & \deltaValue{+4.9} & \deltaValue{+0.3} & \deltaValue{+0.0} & \deltaValue{+0.5} & \deltaValue{+0.1} & \deltaValue{-0.8} & \deltaValue{+0.2} & \deltaValue{+1.7} \\
Gaussian Noise & \deltaValue{+8.5} & \deltaValue{+2.9} & \deltaValue{-2.1} & \deltaValue{-14.3} & \deltaValue{+0.7} & \deltaValue{+0.3} & \deltaValue{-0.5} & \deltaValue{+7.9} & \deltaValue{+8.4} & \deltaValue{+26.6} & \deltaValue{+1.6} & \deltaValue{+3.2} & \deltaValue{-1.1} & \deltaValue{+0.0} & \deltaValue{+0.3} & \deltaValue{+0.3} & \deltaValue{-0.5} & \deltaValue{+0.1} & \deltaValue{+2.3} \\
High Pass Filter & \deltaValue{+1.7} & \deltaValue{-2.3} & \deltaValue{+3.0} & \deltaValue{+0.3} & \deltaValue{-1.9} & \deltaValue{-0.4} & \deltaValue{-2.2} & \deltaValue{+4.4} & \deltaValue{+5.6} & \deltaValue{+1.5} & \deltaValue{+35.2} & \deltaValue{+1.3} & \deltaValue{+1.5} & \deltaValue{-0.2} & \deltaValue{-0.6} & \deltaValue{-0.4} & \deltaValue{+0.2} & \deltaValue{-1.7} & \deltaValue{+2.5} \\
Low Pass Filter & \deltaValue{-0.6} & \deltaValue{+1.1} & \deltaValue{-4.0} & \deltaValue{-13.3} & \deltaValue{+0.5} & \deltaValue{-2.2} & \deltaValue{-0.5} & \deltaValue{+2.4} & \deltaValue{+1.9} & \deltaValue{+7.8} & \deltaValue{+0.6} & \deltaValue{+9.9} & \deltaValue{-0.5} & \deltaValue{+0.0} & \deltaValue{+0.0} & \deltaValue{+0.2} & \deltaValue{-0.8} & \deltaValue{-0.1} & \deltaValue{+0.1} \\
MP3 Compression & \deltaValue{-1.0} & \deltaValue{+0.1} & \deltaValue{-4.9} & \deltaValue{-10.0} & \deltaValue{-0.7} & \deltaValue{-0.9} & \deltaValue{-1.5} & \deltaValue{+3.0} & \deltaValue{+3.1} & \deltaValue{+3.0} & \deltaValue{-0.1} & \deltaValue{+0.6} & \deltaValue{+5.2} & \deltaValue{-0.2} & \deltaValue{-0.1} & \deltaValue{-0.4} & \deltaValue{-0.6} & \deltaValue{-0.6} & \deltaValue{-0.3} \\
Pitch Shift & \deltaValue{+0.5} & \deltaValue{+0.1} & \deltaValue{+4.0} & \deltaValue{-3.5} & \deltaValue{-0.8} & \deltaValue{+5.6} & \deltaValue{-0.8} & \deltaValue{+0.4} & \deltaValue{-2.1} & \deltaValue{+0.8} & \deltaValue{+0.6} & \deltaValue{+1.0} & \deltaValue{-0.5} & \deltaValue{+25.3} & \deltaValue{-0.1} & \deltaValue{-0.2} & \deltaValue{+27.9} & \deltaValue{-0.4} & \deltaValue{+3.2} \\
Reverb & \deltaValue{+1.4} & \deltaValue{-2.0} & \deltaValue{+1.3} & \deltaValue{-8.0} & \deltaValue{-2.2} & \deltaValue{+0.6} & \deltaValue{-2.9} & \deltaValue{+2.0} & \deltaValue{+0.6} & \deltaValue{+4.8} & \deltaValue{+1.8} & \deltaValue{+0.6} & \deltaValue{-1.5} & \deltaValue{+0.0} & \deltaValue{-1.2} & \deltaValue{-2.2} & \deltaValue{-0.8} & \deltaValue{-1.9} & \deltaValue{-0.5} \\
Silence Injection & \deltaValue{-2.7} & \deltaValue{-1.1} & \deltaValue{-1.2} & \deltaValue{-7.3} & \deltaValue{-1.6} & \deltaValue{-1.6} & \deltaValue{-1.3} & \deltaValue{-1.0} & \deltaValue{-1.1} & \deltaValue{+2.0} & \deltaValue{+0.5} & \deltaValue{+1.8} & \deltaValue{+0.5} & \deltaValue{+0.1} & \deltaValue{-0.4} & \deltaValue{-0.1} & \deltaValue{-0.3} & \deltaValue{-0.4} & \deltaValue{-0.8} \\
Time Stretch & \deltaValue{-0.7} & \deltaValue{-1.3} & \deltaValue{+5.9} & \deltaValue{-9.2} & \deltaValue{-0.5} & \deltaValue{+3.6} & \deltaValue{-1.0} & \deltaValue{-3.4} & \deltaValue{-4.1} & \deltaValue{+0.7} & \deltaValue{+1.1} & \deltaValue{+0.4} & \deltaValue{-0.7} & \deltaValue{+27.0} & \deltaValue{-0.2} & \deltaValue{-0.2} & \deltaValue{+30.7} & \deltaValue{-0.3} & \deltaValue{+2.7} \\ \hline
\textbf{No Defense} & 77.9 & 90.2 & 78.2 & 80.5 & 97.1 & 53.5 & 99.3 & 82.7 & 80.6 & 55.5 & 50.0 & 83.9 & 92.2 & 56.0 & 99.2 & 98.7 & 52.7 & 99.4 & 79.3 \\
\textbf{Retrain All} & \deltaValue{+14.5} & \deltaValue{+6.0} & \deltaValue{+19.5} & \deltaValue{+18.9} & \deltaValue{+1.0} & \deltaValue{+42.5} & \deltaValue{+0.0} & \deltaValue{+14.2} & \deltaValue{+15.5} & \deltaValue{+23.4} & \deltaValue{+30.8} & \deltaValue{+11.1} & \deltaValue{+4.9} & \deltaValue{+28.4} & \deltaValue{+0.7} & \deltaValue{-0.1} & \deltaValue{+32.0} & \deltaValue{+0.2} & \deltaValue{+14.6} \\
 \textbf{Retrain Selected}& \deltaValue{+14.5} & \deltaValue{+4.9} & \deltaValue{+18.7} & \deltaValue{+17.0} & \deltaValue{+0.1} & \deltaValue{+42.6} & \deltaValue{-0.9} & \deltaValue{+10.7} & \deltaValue{+12.0} & \deltaValue{+23.8} & \deltaValue{+33.8} & \deltaValue{+2.7} & \deltaValue{+4.8} & \deltaValue{+23.7} & \deltaValue{+0.3} & \deltaValue{-0.2} & \deltaValue{+33.1} & \deltaValue{-0.6} & \deltaValue{+13.4} \\
\bottomrule
\end{tabular}

    }
    \vspace{0.1cm}
    \caption{Performance of an adaptive defender over the MLAAD dataset. The defender's system, a W2V2 model, is retrained  on the complete DeePen dataset (`Retrain All'), or a subset of the attacks. Performance of baseline system, without retraining on DeePen is shown as `No Defense'. Results are averaged over five independent trials, and denote the average accuracy modification relative to the `No Defense' scenario.}
    \label{tab:mlaad_adaptive_delta}
\end{table*}

\begin{table*}[]
    \centering
    \resizebox{.99\textwidth}{!}{
    \begin{tabular}{llllllllllllllllllll}
\toprule
& attack $\rightarrow$ & \rotatebox{90}{Add Background Music} & \rotatebox{90}{Add Background Noise} & \rotatebox{90}{Amplitude Modulation} & \rotatebox{90}{Autotune} & \rotatebox{90}{Bit Depth Change} & \rotatebox{90}{Echo} & \rotatebox{90}{Equalization} & \rotatebox{90}{Freq Minus} & \rotatebox{90}{Freq Plus} & \rotatebox{90}{Gaussian Noise} & \rotatebox{90}{High Pass Filter} & \rotatebox{90}{Low Pass Filter} & \rotatebox{90}{MP3 Compression} & \rotatebox{90}{Pitch Shift} & \rotatebox{90}{Reverb} & \rotatebox{90}{Silence Injection} & \rotatebox{90}{Time Stretch} & \rotatebox{90}{\textbf{No Attack}} \\
label $\downarrow$ & adaptive defense $\downarrow$ &  &  &  &  &  &  &  &  &  &  &  &  &  &  &  &  &  &  \\
\midrule
\multirow[]{20}{*}{\rotatebox{90}{BONA-FIDE}}& Add Background Music & \deltaValue{+18.2} & \deltaValue{+13.4} & \deltaValue{+2.4} & \deltaValue{-9.8} & \deltaValue{+2.2} & \deltaValue{+10.0} & \deltaValue{-0.8} & \deltaValue{-1.6} & \deltaValue{+1.8} & \deltaValue{+15.8} & \deltaValue{+24.6} & \deltaValue{-0.4} & \deltaValue{+0.0} & \deltaValue{+0.4} & \deltaValue{+0.2} & \deltaValue{+0.6} & \deltaValue{+0.0} & \deltaValue{+0.4} \\
& Add Background Noise & \deltaValue{+21.2} & \deltaValue{+17.8} & \deltaValue{+7.2} & \deltaValue{-0.8} & \deltaValue{+2.8} & \deltaValue{+6.4} & \deltaValue{+0.0} & \deltaValue{+6.2} & \deltaValue{+8.2} & \deltaValue{+46.6} & \deltaValue{+18.8} & \deltaValue{+5.2} & \deltaValue{+3.6} & \deltaValue{+0.0} & \deltaValue{+0.2} & \deltaValue{+0.4} & \deltaValue{+0.0} & \deltaValue{+0.4} \\
 & Amplitude Modulation & \deltaValue{+11.8} & \deltaValue{+12.2} & \deltaValue{+31.8} & \deltaValue{-6.0} & \deltaValue{+0.4} & \deltaValue{+2.4} & \deltaValue{-0.4} & \deltaValue{-2.2} & \deltaValue{-2.6} & \deltaValue{+21.4} & \deltaValue{+13.8} & \deltaValue{+2.6} & \deltaValue{-2.8} & \deltaValue{+0.0} & \deltaValue{+0.2} & \deltaValue{+0.6} & \deltaValue{+0.0} & \deltaValue{-0.2} \\
& Autotune & \deltaValue{+2.6} & \deltaValue{+0.8} & \deltaValue{+6.4} & \deltaValue{+54.8} & \deltaValue{+0.8} & \deltaValue{+3.2} & \deltaValue{+0.0} & \deltaValue{-3.6} & \deltaValue{-1.8} & \deltaValue{+2.6} & \deltaValue{+7.6} & \deltaValue{-0.4} & \deltaValue{-2.6} & \deltaValue{+0.0} & \deltaValue{-0.2} & \deltaValue{+0.4} & \deltaValue{+0.0} & \deltaValue{+0.4} \\
& Bit Depth Change & \deltaValue{+10.2} & \deltaValue{+7.4} & \deltaValue{-4.8} & \deltaValue{-2.0} & \deltaValue{+3.2} & \deltaValue{+8.8} & \deltaValue{-0.6} & \deltaValue{-6.6} & \deltaValue{-3.4} & \deltaValue{+19.8} & \deltaValue{+19.2} & \deltaValue{+2.0} & \deltaValue{-1.6} & \deltaValue{+0.0} & \deltaValue{-0.6} & \deltaValue{-0.6} & \deltaValue{+0.0} & \deltaValue{-0.6} \\
& Echo & \deltaValue{+6.4} & \deltaValue{+6.2} & \deltaValue{+3.6} & \deltaValue{+3.8} & \deltaValue{+1.6} & \deltaValue{+76.0} & \deltaValue{+0.2} & \deltaValue{-1.6} & \deltaValue{+1.0} & \deltaValue{+12.0} & \deltaValue{+26.4} & \deltaValue{+1.2} & \deltaValue{+2.2} & \deltaValue{+0.0} & \deltaValue{+0.2} & \deltaValue{+1.2} & \deltaValue{+0.0} & \deltaValue{+0.2} \\
& Equalization & \deltaValue{+3.2} & \deltaValue{+3.2} & \deltaValue{+3.0} & \deltaValue{-3.0} & \deltaValue{+0.2} & \deltaValue{-2.4} & \deltaValue{-0.8} & \deltaValue{-1.4} & \deltaValue{+0.2} & \deltaValue{+21.0} & \deltaValue{+20.6} & \deltaValue{+1.8} & \deltaValue{-3.8} & \deltaValue{+0.0} & \deltaValue{+0.2} & \deltaValue{+1.0} & \deltaValue{+0.0} & \deltaValue{+0.4} \\
& Freq Minus & \deltaValue{+7.6} & \deltaValue{+8.4} & \deltaValue{+8.4} & \deltaValue{+3.8} & \deltaValue{+1.6} & \deltaValue{+3.4} & \deltaValue{-0.4} & \deltaValue{+6.8} & \deltaValue{+8.2} & \deltaValue{+27.6} & \deltaValue{+10.2} & \deltaValue{+3.6} & \deltaValue{+8.2} & \deltaValue{+0.0} & \deltaValue{+0.2} & \deltaValue{+1.2} & \deltaValue{+0.0} & \deltaValue{+0.4} \\
& Freq Plus & \deltaValue{+14.2} & \deltaValue{+13.0} & \deltaValue{+7.4} & \deltaValue{-3.6} & \deltaValue{+3.0} & \deltaValue{-2.2} & \deltaValue{+0.2} & \deltaValue{+6.2} & \deltaValue{+7.4} & \deltaValue{+40.6} & \deltaValue{+9.6} & \deltaValue{+5.6} & \deltaValue{-1.4} & \deltaValue{+0.0} & \deltaValue{+0.2} & \deltaValue{+0.8} & \deltaValue{+0.0} & \deltaValue{+0.4} \\
& Gaussian Noise & \deltaValue{+5.6} & \deltaValue{+15.2} & \deltaValue{+1.6} & \deltaValue{-5.4} & \deltaValue{+2.4} & \deltaValue{+0.0} & \deltaValue{-0.4} & \deltaValue{+2.4} & \deltaValue{+6.2} & \deltaValue{+44.8} & \deltaValue{+18.6} & \deltaValue{+1.4} & \deltaValue{-3.8} & \deltaValue{+0.0} & \deltaValue{+0.2} & \deltaValue{+1.0} & \deltaValue{+0.0} & \deltaValue{+0.0} \\
& High Pass Filter & \deltaValue{+1.8} & \deltaValue{+5.0} & \deltaValue{+12.4} & \deltaValue{+6.8} & \deltaValue{+2.6} & \deltaValue{-0.2} & \deltaValue{-0.4} & \deltaValue{+3.2} & \deltaValue{+5.8} & \deltaValue{+14.4} & \deltaValue{+94.6} & \deltaValue{+6.8} & \deltaValue{+8.6} & \deltaValue{+0.0} & \deltaValue{+0.2} & \deltaValue{+1.2} & \deltaValue{+0.0} & \deltaValue{+0.4} \\
& Low Pass Filter & \deltaValue{+10.2} & \deltaValue{+11.8} & \deltaValue{-0.6} & \deltaValue{-3.0} & \deltaValue{+1.8} & \deltaValue{-4.2} & \deltaValue{-0.8} & \deltaValue{+0.4} & \deltaValue{+2.4} & \deltaValue{+33.8} & \deltaValue{+9.6} & \deltaValue{+10.8} & \deltaValue{-3.6} & \deltaValue{+0.0} & \deltaValue{+0.2} & \deltaValue{+1.0} & \deltaValue{+0.0} & \deltaValue{+0.4} \\
& MP3 Compression & \deltaValue{-3.0} & \deltaValue{+4.0} & \deltaValue{+2.8} & \deltaValue{+6.2} & \deltaValue{+2.4} & \deltaValue{+3.6} & \deltaValue{-0.4} & \deltaValue{+3.2} & \deltaValue{+3.8} & \deltaValue{+26.4} & \deltaValue{+2.4} & \deltaValue{+5.4} & \deltaValue{+17.8} & \deltaValue{+0.0} & \deltaValue{+0.2} & \deltaValue{+0.4} & \deltaValue{+0.0} & \deltaValue{+0.4} \\
& Pitch Shift & \deltaValue{+11.4} & \deltaValue{+9.0} & \deltaValue{+5.0} & \deltaValue{+0.8} & \deltaValue{+1.0} & \deltaValue{+10.2} & \deltaValue{-0.6} & \deltaValue{-3.0} & \deltaValue{-0.6} & \deltaValue{+4.8} & \deltaValue{+15.8} & \deltaValue{+1.2} & \deltaValue{+10.4} & \deltaValue{+81.6} & \deltaValue{+0.0} & \deltaValue{+0.4} & \deltaValue{+85.6} & \deltaValue{+0.0} \\
& Reverb & \deltaValue{+13.0} & \deltaValue{+10.4} & \deltaValue{+6.6} & \deltaValue{+2.6} & \deltaValue{+0.0} & \deltaValue{+4.8} & \deltaValue{-2.4} & \deltaValue{-3.6} & \deltaValue{-0.8} & \deltaValue{+30.6} & \deltaValue{+9.8} & \deltaValue{+1.8} & \deltaValue{-0.8} & \deltaValue{+0.0} & \deltaValue{-0.4} & \deltaValue{-2.8} & \deltaValue{+0.0} & \deltaValue{-2.0} \\
& Silence Injection & \deltaValue{+1.8} & \deltaValue{+2.2} & \deltaValue{-0.6} & \deltaValue{+3.8} & \deltaValue{-1.6} & \deltaValue{-1.6} & \deltaValue{-0.8} & \deltaValue{-2.0} & \deltaValue{+0.8} & \deltaValue{+5.4} & \deltaValue{+14.0} & \deltaValue{+2.6} & \deltaValue{-2.2} & \deltaValue{+0.0} & \deltaValue{+0.2} & \deltaValue{+0.4} & \deltaValue{+0.0} & \deltaValue{-0.6} \\
& Time Stretch & \deltaValue{+0.4} & \deltaValue{+4.8} & \deltaValue{+5.6} & \deltaValue{-1.4} & \deltaValue{-2.2} & \deltaValue{+12.2} & \deltaValue{-0.6} & \deltaValue{-4.2} & \deltaValue{-2.2} & \deltaValue{+4.8} & \deltaValue{+14.8} & \deltaValue{+1.8} & \deltaValue{+7.6} & \deltaValue{+76.4} & \deltaValue{-0.2} & \deltaValue{+1.2} & \deltaValue{+87.4} & \deltaValue{-0.6} \\ \hline
& \textbf{No Defense} & 75.0 & 77.6 & 64.4 & 33.4 & 96.4 & 20.0 & 99.8 & 91.6 & 91.6 & 51.4 & 2.0 & 87.4 & 80.2 & 11.0 & 99.8 & 98.8 & 3.0 & 99.6 \\
& \textbf{Retrain All} & \deltaValue{+23.4} & \deltaValue{+21.4} & \deltaValue{+33.8} & \deltaValue{+58.4} & \deltaValue{+3.6} & \deltaValue{+76.8} & \deltaValue{+0.2} & \deltaValue{+8.0} & \deltaValue{+8.4} & \deltaValue{+46.2} & \deltaValue{+95.2} & \deltaValue{+12.6} & \deltaValue{+19.4} & \deltaValue{+84.2} & \deltaValue{+0.2} & \deltaValue{+1.2} & \deltaValue{+91.8} & \deltaValue{+0.4} \\
& \textbf{Retrain Selected} & \deltaValue{+21.4} & \deltaValue{+19.6} & \deltaValue{+32.8} & \deltaValue{+57.2} & \deltaValue{+3.4} & \deltaValue{+72.0} & \deltaValue{+0.2} & \deltaValue{+5.2} & \deltaValue{+6.6} & \deltaValue{+45.2} & \deltaValue{+83.0} & \deltaValue{+5.6} & \deltaValue{+19.2} & \deltaValue{+65.8} & \deltaValue{+0.2} & \deltaValue{+1.2} & \deltaValue{+77.8} & \deltaValue{+0.4} \\ \hline \hline
\multirow[]{20}{*}{\rotatebox{90}{SPOOF}}& Add Background Music & \deltaValue{-2.6} & \deltaValue{+1.0} & \deltaValue{+0.0} & \deltaValue{+0.0} & \deltaValue{-0.2} & \deltaValue{-0.6} & \deltaValue{+0.0} & \deltaValue{+1.0} & \deltaValue{+1.6} & \deltaValue{-0.8} & \deltaValue{-16.4} & \deltaValue{+2.4} & \deltaValue{+0.0} & \deltaValue{-0.4} & \deltaValue{+0.6} & \deltaValue{+1.2} & \deltaValue{-0.2} & \deltaValue{+0.0} \\
& Add Background Noise & \deltaValue{-6.2} & \deltaValue{-1.4} & \deltaValue{+0.0} & \deltaValue{+0.0} & \deltaValue{-0.2} & \deltaValue{+0.2} & \deltaValue{+0.4} & \deltaValue{-0.8} & \deltaValue{-2.2} & \deltaValue{-15.6} & \deltaValue{-5.8} & \deltaValue{-3.2} & \deltaValue{+0.0} & \deltaValue{+0.0} & \deltaValue{-0.2} & \deltaValue{+0.4} & \deltaValue{+0.0} & \deltaValue{+0.0} \\
& Amplitude Modulation & \deltaValue{-9.2} & \deltaValue{-2.6} & \deltaValue{-2.4} & \deltaValue{+0.0} & \deltaValue{-0.2} & \deltaValue{+0.0} & \deltaValue{-0.2} & \deltaValue{+0.2} & \deltaValue{+0.0} & \deltaValue{-5.2} & \deltaValue{-7.0} & \deltaValue{-1.0} & \deltaValue{+0.0} & \deltaValue{+0.4} & \deltaValue{+0.2} & \deltaValue{+0.6} & \deltaValue{+0.0} & \deltaValue{+0.4} \\
& Autotune & \deltaValue{-4.8} & \deltaValue{-2.2} & \deltaValue{-0.2} & \deltaValue{-1.0} & \deltaValue{-0.4} & \deltaValue{-1.0} & \deltaValue{-0.2} & \deltaValue{-0.4} & \deltaValue{-0.8} & \deltaValue{-2.6} & \deltaValue{-0.4} & \deltaValue{+0.6} & \deltaValue{+0.0} & \deltaValue{+0.0} & \deltaValue{-1.0} & \deltaValue{+0.2} & \deltaValue{+0.0} & \deltaValue{-0.6} \\
& Bit Depth Change & \deltaValue{-5.8} & \deltaValue{-1.4} & \deltaValue{+0.0} & \deltaValue{+0.0} & \deltaValue{-0.2} & \deltaValue{-0.8} & \deltaValue{-0.2} & \deltaValue{-0.4} & \deltaValue{+0.2} & \deltaValue{-4.8} & \deltaValue{-6.6} & \deltaValue{-1.6} & \deltaValue{-0.2} & \deltaValue{+0.0} & \deltaValue{+0.2} & \deltaValue{+0.6} & \deltaValue{+0.0} & \deltaValue{+0.4} \\
& Echo & \deltaValue{-8.8} & \deltaValue{-2.2} & \deltaValue{-0.8} & \deltaValue{+0.0} & \deltaValue{-0.2} & \deltaValue{-10.8} & \deltaValue{-0.2} & \deltaValue{-1.6} & \deltaValue{-0.8} & \deltaValue{-2.0} & \deltaValue{-16.8} & \deltaValue{-1.6} & \deltaValue{+0.0} & \deltaValue{+0.2} & \deltaValue{-1.0} & \deltaValue{+0.0} & \deltaValue{-0.2} & \deltaValue{+0.0} \\
& Equalization & \deltaValue{+1.4} & \deltaValue{+0.4} & \deltaValue{-0.4} & \deltaValue{+0.0} & \deltaValue{+0.0} & \deltaValue{+0.2} & \deltaValue{+0.8} & \deltaValue{+0.8} & \deltaValue{+1.0} & \deltaValue{-3.2} & \deltaValue{-7.6} & \deltaValue{+2.0} & \deltaValue{+0.0} & \deltaValue{+0.2} & \deltaValue{+0.6} & \deltaValue{+0.2} & \deltaValue{+0.0} & \deltaValue{+0.2} \\
& Freq Minus & \deltaValue{-3.8} & \deltaValue{-0.8} & \deltaValue{+0.0} & \deltaValue{+0.0} & \deltaValue{-0.2} & \deltaValue{+0.2} & \deltaValue{+0.4} & \deltaValue{-0.4} & \deltaValue{+0.6} & \deltaValue{-2.2} & \deltaValue{-2.4} & \deltaValue{-2.6} & \deltaValue{-0.2} & \deltaValue{+0.0} & \deltaValue{+0.2} & \deltaValue{-0.4} & \deltaValue{+0.0} & \deltaValue{-0.2} \\
& Freq Plus & \deltaValue{-6.8} & \deltaValue{-1.0} & \deltaValue{-0.2} & \deltaValue{+0.0} & \deltaValue{+0.0} & \deltaValue{+0.2} & \deltaValue{+0.4} & \deltaValue{+0.0} & \deltaValue{-0.2} & \deltaValue{-4.0} & \deltaValue{-5.2} & \deltaValue{-3.0} & \deltaValue{+0.0} & \deltaValue{+0.2} & \deltaValue{+1.0} & \deltaValue{+0.8} & \deltaValue{+0.0} & \deltaValue{+0.4} \\
& Gaussian Noise & \deltaValue{+3.8} & \deltaValue{+0.4} & \deltaValue{-0.2} & \deltaValue{+0.0} & \deltaValue{-0.4} & \deltaValue{+0.2} & \deltaValue{+0.8} & \deltaValue{+0.2} & \deltaValue{+1.0} & \deltaValue{-1.6} & \deltaValue{-1.0} & \deltaValue{+2.0} & \deltaValue{+0.0} & \deltaValue{+0.2} & \deltaValue{+0.8} & \deltaValue{+1.0} & \deltaValue{+0.0} & \deltaValue{+0.0} \\
& High Pass Filter & \deltaValue{-2.0} & \deltaValue{-1.2} & \deltaValue{-0.8} & \deltaValue{+0.0} & \deltaValue{+0.0} & \deltaValue{+0.2} & \deltaValue{-0.2} & \deltaValue{-2.0} & \deltaValue{-0.6} & \deltaValue{-2.4} & \deltaValue{-22.0} & \deltaValue{-6.0} & \deltaValue{-0.2} & \deltaValue{+0.0} & \deltaValue{-0.8} & \deltaValue{-0.2} & \deltaValue{+0.0} & \deltaValue{-0.4} \\
& Low Pass Filter & \deltaValue{-3.2} & \deltaValue{-2.0} & \deltaValue{+0.0} & \deltaValue{+0.0} & \deltaValue{+0.0} & \deltaValue{+0.0} & \deltaValue{+0.4} & \deltaValue{-0.4} & \deltaValue{+0.8} & \deltaValue{-11.0} & \deltaValue{-2.8} & \deltaValue{-1.2} & \deltaValue{+0.0} & \deltaValue{+0.0} & \deltaValue{+0.6} & \deltaValue{+0.2} & \deltaValue{+0.0} & \deltaValue{+0.0} \\
& MP3 Compression & \deltaValue{-0.6} & \deltaValue{-2.0} & \deltaValue{-0.8} & \deltaValue{+0.0} & \deltaValue{-0.6} & \deltaValue{-0.2} & \deltaValue{-0.6} & \deltaValue{-2.2} & \deltaValue{-2.6} & \deltaValue{-6.8} & \deltaValue{+0.0} & \deltaValue{-7.2} & \deltaValue{-2.6} & \deltaValue{+0.0} & \deltaValue{-0.6} & \deltaValue{+0.0} & \deltaValue{+0.0} & \deltaValue{-0.4} \\
& Pitch Shift & \deltaValue{-5.2} & \deltaValue{-1.0} & \deltaValue{-0.2} & \deltaValue{+0.0} & \deltaValue{+0.0} & \deltaValue{-0.6} & \deltaValue{+0.0} & \deltaValue{-0.4} & \deltaValue{+0.4} & \deltaValue{-0.6} & \deltaValue{-5.0} & \deltaValue{-0.2} & \deltaValue{+0.0} & \deltaValue{-35.4} & \deltaValue{+0.8} & \deltaValue{+0.8} & \deltaValue{-27.4} & \deltaValue{+0.0} \\
& Reverb & \deltaValue{-6.6} & \deltaValue{-3.2} & \deltaValue{-1.0} & \deltaValue{+0.0} & \deltaValue{-0.4} & \deltaValue{+0.2} & \deltaValue{-0.2} & \deltaValue{-1.2} & \deltaValue{-1.4} & \deltaValue{-17.2} & \deltaValue{-2.2} & \deltaValue{-3.2} & \deltaValue{-0.2} & \deltaValue{+0.2} & \deltaValue{-0.6} & \deltaValue{+0.4} & \deltaValue{+0.0} & \deltaValue{-0.2} \\
& Silence Injection & \deltaValue{-3.8} & \deltaValue{-1.4} & \deltaValue{-0.2} & \deltaValue{+0.0} & \deltaValue{-0.2} & \deltaValue{-0.8} & \deltaValue{+0.2} & \deltaValue{-1.8} & \deltaValue{-0.8} & \deltaValue{-1.0} & \deltaValue{-3.0} & \deltaValue{-1.4} & \deltaValue{+0.0} & \deltaValue{+0.2} & \deltaValue{+0.4} & \deltaValue{+0.2} & \deltaValue{+0.0} & \deltaValue{+0.0} \\
& Time Stretch & \deltaValue{+0.0} & \deltaValue{+0.0} & \deltaValue{+0.0} & \deltaValue{+0.0} & \deltaValue{+0.0} & \deltaValue{-0.2} & \deltaValue{+0.2} & \deltaValue{+0.2} & \deltaValue{+1.2} & \deltaValue{-0.6} & \deltaValue{-3.2} & \deltaValue{-0.2} & \deltaValue{+0.0} & \deltaValue{-28.2} & \deltaValue{+0.4} & \deltaValue{+0.8} & \deltaValue{-20.0} & \deltaValue{+0.0} \\ \hline
& \textbf{No Defense} & 93.2 & 97.0 & 100.0 & 100.0 & 100.0 & 99.8 & 98.8 & 98.8 & 98.0 & 98.0 & 99.6 & 92.2 & 100.0 & 99.0 & 98.6 & 97.8 & 100.0 & 99.2 \\
& \textbf{Retrain All} & \deltaValue{-5.0} & \deltaValue{-2.2} & \deltaValue{-3.6} & \deltaValue{-0.6} & \deltaValue{-0.2} & \deltaValue{-9.4} & \deltaValue{-0.6} & \deltaValue{-0.2} & \deltaValue{+0.6} & \deltaValue{-3.4} & \deltaValue{-25.4} & \deltaValue{-5.8} & \deltaValue{-3.0} & \deltaValue{-33.8} & \deltaValue{-1.8} & \deltaValue{-0.2} & \deltaValue{-25.8} & \deltaValue{-1.0} \\
& \textbf{Retrain Selected} & \deltaValue{-0.6} & \deltaValue{-1.8} & \deltaValue{-2.2} & \deltaValue{-1.6} & \deltaValue{+0.0} & \deltaValue{-5.8} & \deltaValue{+0.2} & \deltaValue{+0.2} & \deltaValue{+1.4} & \deltaValue{-0.8} & \deltaValue{-10.4} & \deltaValue{-5.6} & \deltaValue{-1.8} & \deltaValue{-18.0} & \deltaValue{-1.0} & \deltaValue{+1.0} & \deltaValue{-14.8} & \deltaValue{-1.0} \\

\bottomrule
\end{tabular}

    }
    \vspace{0.1cm}
    \caption{ASVspoof 2019 performance of DeePen for an adaptive defender, who trains either on the complete DeePen dataset (``Retrain All''), or some attacks (``Retrain Selected''). 
    Performance of baseline system, without retraining on DeePen, shown as `No Defense'. Results averaged over five independent trials, with average accuracy relative to the `no defense' scenario shown.
    Red highlights indicate a deterioration of the system in percent absolute over the baseline by more than $5\%$, while blue highlights indicate an improvement by more than $5\%$. 
    }
    \label{tab:asv19_adaptive_carrier_delta}
\end{table*}

\end{document}